\documentclass{ws-procs9x6}
\usepackage{graphicx}

\begin{document}
\newcommand{\be}{\begin{equation}}
\newcommand{\ee}{\end{equation}}
\newcommand{\bea}{\begin{eqnarray}}
\newcommand{\eea}{\end{eqnarray}}
\newcommand{\ba}{\begin{array}}
\newcommand{\ea}{\end{array}}
\newcommand{\bc}{\begin{center}}
\newcommand{\ec}{\end{center}}
\renewcommand{\slash}{\!\!\!\!/\,}
\newcommand{\sslash}{\!\!\!/\,}
\newcommand{\Dslash}{D\hspace*{-0.23cm}/\,}

\title{Phases of QCD}

\author{Thomas Sch\"afer}

\address{Department of Physics\\
     North Carolina State University\\
     Raleigh, NC 27695}

\maketitle

\abstracts{
In these lectures we provide an introduction to the phase structure
of QCD. We begin with a brief discussion of QCD, the symmetries 
of QCD, and what we mean by a ``phase of QCD''. In the main 
part of the lectures we discuss the phase diagram of QCD as 
a function of the temperature and the baryon density. We 
focus, in particular, on the high temperature plasma phase, 
the low temperature and low density nuclear phase, and the 
high density color superconducting phases. }

\section{Introduction}
\label{sec_intro}

 In these lectures we wish to provide an introduction to the 
phase structure of QCD. The phase of QCD that we live in is 
characterized by the permanent confinement of quarks, and the 
existence of a large mass gap. There are several reasons for 
trying to understand whether other phases of QCD exist, and 
what conditions are required in order to observe these phases:

1) Other phases of QCD exist in the universe: The universe
started out in a hot and dense phase. It expanded and 
cooled and about $10^{-5}$ sec after the big bang it 
passed through a transition from a quark gluon plasma 
to a hadronic phase. Even today, extreme conditions
exist in the universe. In supernova explosions matter is 
heated to several tens of MeV, sufficient to evaporate 
ordinary nuclei. The compact remnants have central densities 
several times larger than the saturation density of 
nuclear matter. 

2) Exploring the entire phase diagram help us to understand
the phase that we live in: The structure of hadrons and their 
interactions are determined by the symmetries of the QCD vacuum.
Studying the phase diagram of QCD allows us to understand
the possible ways in which the symmetries of QCD can be 
realized. 

3) QCD simplifies in extreme environments: At scales
relevant to hadrons QCD is strongly coupled and we have
to rely on numerical simulations in order to test predictions
of QCD. In the case of large temperature or large baryon
density there is a large external scale in the problem.
Asymptotic freedom implies that the bulk of the system is
governed by weak coupling. As a result, we can study
QCD matter in a regime where quarks and gluons are indeed
the correct degrees of freedom.

 In these lectures we will give a general introduction into 
the physics of the QCD phase diagram. There are several excellent 
text books and reviews articles that provide a much more detailed
discussion of QCD and hadronic matter at finite temperature and density 
\cite{Shuryak:1988,Kogut:2004su,Rischke:2003mt}. We also recommend
more specialized texts on field theory at finite temperature
\cite{Kapusta:1989,LeBellac:1996,Kraemmer:2003gd} and density 
\cite{Fetter:1971,Abrikosov:1963}, as well as reviews on the 
phase structure of dense matter 
\cite{Rajagopal:2000wf,Alford:2001dt,Schafer:2003vz}
and on color superconductivity 
\cite{Buballa:2003qv,Ren:2004nn,Huang:2004ik,Shovkovy:2004me}. 
In this write-up we will not try to give a summary of the 
experimental program at RHIC and the SPS. A useful reference 
is the series of white papers that was recently published by the 
RHIC collaborations \cite{rhic:2005}. We will also not review 
implications of the phase structure of QCD for the structure 
of compact stars or observational constraints on the behavior
of dense matter \cite{Alford:2001dt,Nardulli:2002ma,Reddy:2002ri}.

\section{QCD and symmetries}
\label{sec_qcd}
\subsection{Introduction}
\label{sec_qcd_intro}

 We begin with a brief review of QCD and the symmetries
of QCD. The elementary degrees of freedom are quark fields 
$\psi^a_{\alpha,f}$ and gluons $A_\mu^a$. Here, $a$ is 
color index that transforms in the fundamental representation
for fermions and in the adjoint representation for gluons. 
Also, $f$ labels the quark flavors $u,d,s,c,b,t$. In practice, 
we will focus on the three light flavors up, down and strange.
The QCD lagrangian is 
\be
\label{l_qcd}
 {\mathcal L } = \sum_f^{N_f} \bar{\psi}_f ( i\Dslash - m_f) \psi_f
  - \frac{1}{4} G_{\mu\nu}^a G_{\mu\nu}^a,
\ee
where the field strength tensor is defined by 
\be
 G_{\mu\nu}^a = \partial_\mu A_\nu^a - \partial_\nu A_\mu^a
  + gf^{abc} A_\mu^b A_\nu^c,
\ee
and the covariant derivative acting on quark fields is 
\be
 i\Dslash \psi = \gamma^\mu \left(
 i\partial_\mu + g A_\mu^a \frac{\lambda^a}{2}\right) \psi.
\ee
QCD has a number of interesting properties. Most remarkably, 
even though QCD accounts for the rich phenomenology of hadronic
and nuclear physics, it is an essentially parameter free 
theory. As a first approximation, the masses of the light
quarks $u,d,s$ are too small to be important, while the masses 
of the heavy quarks $c,b,t$ are too heavy. If we set 
the masses of the light quarks to zero and take the masses
of the heavy quarks to be infinite then the only parameter 
in the QCD lagrangian is the coupling constant, $g$. Once
quantum corrections are taken into account $g$ becomes a 
function of the scale at which it is measured. Gross, Wilczek 
and Politzer showed that \cite{Gross:1973id,Politzer:1973fx}
\be 
\label{as_fr}
 g^2(q^2) = \frac{16\pi^2}{b\log(q^2/\Lambda^2_{QCD})}, 
\hspace{0.3cm} 
b =\frac{11N_c}{3}-\frac{2N_f}{3}.
\ee
If the scale $q^2$ is large then the coupling is small, but 
in the infrared the coupling becomes large. This is the famous
phenomenon of asymptotic freedom. Since the coupling 
depends on the scale the dimensionless parameter $g$ is 
traded for a dimensionful scale parameter $\Lambda_{QCD}$. 
In essence, $\Lambda_{QCD}$ is the scale at which the 
theory becomes non-perturbative. 

 Since $\Lambda_{QCD}$ is the only dimensionful quantity
in QCD ($m_q=0$) it is not really a parameter of QCD, but
reflects our choice of units. In standard units, $\Lambda_{QCD} 
\simeq 200\,{\rm MeV} \simeq 1\,{\rm fm}^{-1}$. Note that 
hadrons indeed have sizes $r\sim\Lambda_{QCD}^{-1}$. 

 Another important feature of the QCD lagrangian are 
its symmetries. First of all, the lagrangian is invariant 
under local gauge transformations $U(x)\in SU(3)_c$
\be 
\psi(x) \to U(x)\psi(x),\hspace{1cm}
A_\mu(x) \to U(x)A_\mu U^\dagger (x)
 + iU(x)\partial_\mu U^\dagger(x),
\ee
where $A_\mu= A_\mu^a(\lambda^a/2)$. While the gauge symmetry 
is intimately connected with the dynamics of QCD we observe that
the interactions are completely independent of flavor. If the 
masses of the quarks are equal, $m_u=m_d=m_s$, then the theory 
is invariant under arbitrary flavor rotations of the quark fields 
\be
 \psi_f\to V_{fg}\psi_g,
\ee
where 
$V\in SU(3)$. This is the well known flavor (isospin)
symmetry of the strong interactions. If the quark masses 
are not just equal, but equal to zero, then the flavor 
symmetry is enlarged. This can be seen by defining left 
and right-handed fields
\be
  \psi_{L,R} = \frac{1}{2} (1\pm \gamma_5) \psi .
\ee
In terms of $L/R$ fields the fermionic lagrangian is
\be
 {\mathcal L} =  \bar{\psi}_L (i\Dslash) \psi_L
    +\bar{\psi}_R (i\Dslash) \psi_R + 
   \bar{\psi}_L M \psi_R + \bar{\psi}_R M\psi_L ,
\ee
where $M = {\rm diag}(m_u,m_d,m_s)$. We observe that if 
quarks are massless, $m_u=m_d=m_s=0$, then there is no 
coupling between left and right handed fields. As a 
consequence, the lagrangian is invariant under independent
flavor transformations of the left and right handed fields.
\be
 \psi_{L,f}\to L_{fg}\psi_{L,g}, \hspace{1cm}
 \psi_{R,f}\to R_{fg}\psi_{R,g},
\ee
where $(L,R)\in SU(3)_L\times SU(3)_R$. In the real world, 
of course, the masses of the up, down and strange quarks 
are not zero. Nevertheless, since $m_u,m_d\ll m_s < \Lambda_{QCD}$ 
QCD has an approximate chiral symmetry. 

 Finally, we observe that the QCD lagrangian has two 
$U(1)$ symmetries,
\bea
U(1)_B: \hspace{1cm}& \psi_L\to e^{i\phi}\psi_L, \hspace{1cm}& 
     \psi_R\to e^{i\phi}\psi_R , \\
U(1)_A: \hspace{1cm}& \psi_L\to e^{i\alpha}\psi_L,\hspace{1cm} & 
     \psi_R\to e^{-i\alpha}\psi_R .
\eea
The $U(1)_B$ symmetry is exact even if the quarks are not 
massless. The axial $U(1)_A$ symmetry is exact at the classical
level but it is broken in the quantum theory. This phenomenon is 
referred to as an anomaly. The divergence of the $U(1)_A$ current 
is given by
\be 
\partial^\mu j_\mu^5 = \frac{N_f g^2}{16\pi^2}
 G^a_{\mu\nu}\tilde{G}^a_{\mu\nu},
\ee
where $\tilde{G}^a_{\mu\nu}=\epsilon_{\mu\nu\alpha\beta}
G^a_{\alpha\beta}/2$ is the dual field strength tensor. 

\subsection{Phases of QCD}
\label{sec_qcd_phases}

 The phases of QCD are related to the different ways in which 
the symmetries of QCD can be realized in nature. We first
consider the local gauge symmetry. There are three possible
realizations of a local symmetry:

 1) Coulomb Phase: In a Coulomb phase the gauge symmetry is 
unbroken, the gauge bosons are massless and mediate long 
range forces. In particular, the potential between two heavy 
charges is a Coulomb potential, $V(r)\sim 1/r$. 

 2) Higgs Phase: In a Higgs phase the gauge symmetry is 
spontaneously broken and the gauge bosons acquire a mass. 
As a consequence, the potential between two heavy charges
is a Yukawa potential, $V(r)\sim \exp(-mr)/r$. We should
note that local gauge symmetry is related to the fact
that we are using redundant variables (the four-component
vector potential $A_\mu$ describes the two polarization states
of a massless vector boson), and that therefore a local symmetry 
cannot really be broken (Elitzur's theorem \cite{Elitzur:1975im}). 
We will discuss the exact meaning of ``spontaneous gauge 
symmetry breaking'' in Sect.~\ref{sec_lg} below. 

 3) Confinement: In a confined phase all the physical 
excitations are singlets under the gauge group. Confinement 
can be strictly defined only in theories that do not have
light fields in the fundamental representation. In that 
case, confinement implies that the potential between
two heavy charges rises linearly, $V(r)\sim kr$. This is 
called a string potential. If there are light fields in 
the fundamental representation, as in QCD with light quarks,  
then the string can break and the potential levels off. 

 It is interesting to note that all three realizations
of gauge symmetry play a role in the standard model. The 
$U(1)$ of electromagnetism is in a Coulomb phase, the 
$SU(2)$ is realized in a Higgs phase, and the $SU(3)$ of
color is confined. Also, as we shall see in these lectures,
there are phases of QCD in which the color symmetry is 
not confined but realized in a Higgs or Coulomb phase. 

 Different phases of matter, like liquid vs solid, superfluid 
vs normal, are related to the realization of global symmetries. 
In QCD at zero baryon density spacetime symmetries as well as 
$U(1)$ symmetries cannot be broken \cite{Vafa:tf,Vafa:1984xg}. 
This means that phases of QCD matter are governed by the 
realization of the chiral $SU(3)_L\times SU(3)_R$ symmetry. 
If the baryon density is not zero both space-time and $U(1)$
symmetries can break and the phase diagram is much richer. 

\subsection{The QCD vacuum}
\label{sec_qcd_vac}

 In the QCD ground state at zero temperature and density 
chiral symmetry is spontaneously broken by a quark-anti-quark 
condensate $\langle \bar\psi\psi\rangle$. We can view the chiral 
condensate as a matrix in flavor space. In the QCD vacuum
\be
\langle\bar\psi_L^f\psi^g_R\rangle = 
\langle\bar\psi_R^f\psi^g_L\rangle \simeq
 -\delta^{fg} (230\,{\rm MeV})^3 ,
\ee
which implies that chiral symmetry is spontaneously broken according 
to $SU(3)_L\times SU(3)_R\to SU(3)_V$. The $SU(3)_V$ flavor symmetry
is broken explicitly by the difference between the masses of the 
up, down and strange quark. Since $m_s\gg m_u,m_d$ the $SU(2)$
isospin symmetry is a much better symmetry than $SU(3)$ flavor
symmetry. 
 
 Chiral symmetry breaking has important consequences for the 
dynamics of QCD at low energy. Goldstone's theorem implies that 
the breaking of $SU(3)_L\times SU(3)_R\to SU(3)_V$ is associated 
with the appearance of an octet of (approximately) massless 
pseudoscalar Goldstone bosons. Chiral symmetry places important 
restrictions on the interaction of the Goldstone bosons. These 
constraints are most easily obtained from the low energy effective 
chiral lagrangian. The transformations properties of the chiral
field $\Sigma$ follow from the structure of the chiral order 
parameter, 
\be
 \Sigma \to L\Sigma R^\dagger, \hspace{1cm}
 \Sigma^\dagger \to R\Sigma^\dagger L^\dagger,
\ee
for $(L,R)\in SU(3)_L\times SU(3)_R$. In the vacuum we can take
$\langle\Sigma\rangle = 1$. Goldstone modes are fluctuations of 
the order parameter in the coset space $SU(3)_L\times SU(3)_R/
SU(3)_V$. They are parameterized by unitary matrices $\Sigma = 
\exp(i\lambda^a\phi^a/f_\pi)$ where $\lambda^a$ are the Gell-Mann
matrices and $f_\pi=93$ MeV is the pion decay constant. At low 
energy the effective lagrangian for $\Sigma$ can be organized as 
an expansion in the number of derivatives. At leading order in 
$(\partial/f_\pi)$ there is only one structure which is consistent
with chiral symmetry, Lorentz invariance and C,P,T. This is the 
lagrangian of the non-linear sigma model
\be
\label{l_chpt}
{\mathcal L} = \frac{f_\pi^2}{4} {\rm Tr}\left[
 \partial_\mu\Sigma\partial^\mu\Sigma^\dagger\right] 
+ \ldots. 
\ee
In order to show that the parameter $f_\pi$ is related to the
pion decay amplitude we have to gauge the non-linear sigma model. 
This is achieved by introducing the gauge covariant derivative 
$\nabla_\mu\Sigma = \partial_\mu\Sigma+ig_w W_\mu\Sigma$ where 
$W_\mu$ is the charged weak gauge boson and $g_w$ is the weak
coupling constant. The gauged non-linear sigma model gives a
pion-$W$ boson interaction ${\mathcal L}=g_w f_\pi W^\pm_\mu 
\partial^\mu \pi^\mp$ which agrees with the standard definition
of $f_\pi$ in terms of the pion-weak axial current matrix 
element.

 Expanding $\Sigma$ in powers of the pion, kaon and eta fields 
$\phi^a$ we can derive low energy predictions for Goldstone boson 
scattering. In the pion sector we have
\be
{\mathcal L} =\frac{1}{2}(\partial_\mu\phi^a)^2
+\frac{1}{6f_\pi^2}\left[ (\phi^a\partial_\mu \phi^a)^2
  -(\phi^a)^2(\partial_\mu\phi^b)^2 \right] +
  O\left(\frac{\partial^4}{f_\pi^4}\right),
\ee
which shows that the low energy $\pi\pi$-scattering amplitude 
is completely determined by $f_\pi$. Higher order corrections 
originate from loops and higher order terms in the effective 
lagrangian. 

 In QCD chiral symmetry is explicitly broken by the 
quark mass term $\bar\psi M\psi$, where $M={\rm diag}
(m_u,m_d,m_s)$ is the quark mass matrix. In order to determine
how the quark masses appear in the effective lagrangian it 
is useful to promote the mass matrix to a field which 
transforms as $M\to LMR^\dagger$ under chiral transformations. 
This means that the mass term is chirally invariant and 
explicit breaking only appears when $M$ is replaced by its
vacuum value. There is a unique term in the chiral lagrangian 
which is $SU(3)_L\times SU(3)_R$ invariant and linear in 
$M$. To order $O(\partial^2,M)$ the effective lagrangian is 
\be
\label{l_chpt_m}
{\mathcal L} = \frac{f_\pi^2}{4} {\rm Tr}\left[
 \partial_\mu\Sigma\partial^\mu\Sigma^\dagger\right] 
  +\left[ B {\rm Tr}(M\Sigma^\dagger) + h.c. \right]
+ \ldots. 
\ee
The mass term acts a potential for the chiral field. We 
observe that if the quark masses are real and positive then 
the minimum of the potential is at $\langle\Sigma\rangle = 1$,
as expected. If some of the quark masses are negative unusual
phases of QCD can appear, see \cite{Dashen:1970et}.

 The vacuum energy is $E_{vac}=-2B{\rm Tr}[M]$. Using 
$\langle\bar\psi\psi\rangle = \partial E_{vac}/(\partial 
m)$ we find $\langle\bar\psi\psi\rangle=-2B$. Fluctuations 
around the vacuum value $\Sigma=1$ determine the Goldstone
boson masses. The pion mass satisfies the Gell-Mann-Oaks-Renner 
relation
\be
\label{GMOR}
m_\pi^2 f_\pi^2 = (m_u+m_d)\langle\bar\psi\psi\rangle
\ee
and analogous relations exist for the kaon and eta masses.

\subsection{QCD vacuum for different $N_c$ and $N_f$}
\label{sec_bigpic}

\begin{figure}[t]
\bc\includegraphics[width=11.0cm]{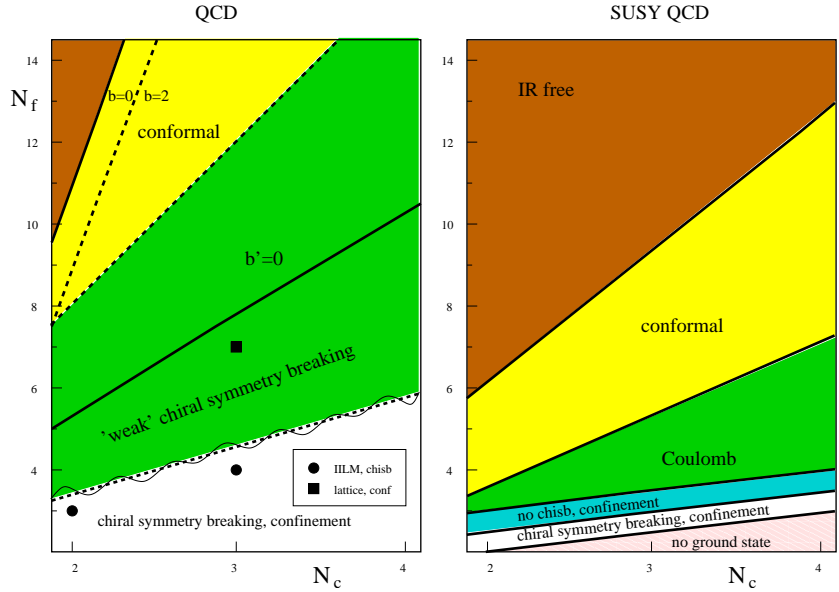}\ec
\caption{\label{fig_bigpic}
Ground state of QCD and SUSY QCD as a function of $N_c$ and 
$N_f$. The symmetry breaking pattern in SUSY QCD was clarified
in a series of papers by Seiberg and collaborators. The phase 
structure of QCD is an educated guess, discussed in more detail 
in the review (Sch\"afer and Shuryak, 1998).}   .
\end{figure}

 QCD is a strongly interacting gauge theory with almost massless
quarks. It seems natural that in such a theory bound states 
of quarks and anti-quarks are formed, that bound states in 
the scalar channel condense, and that chiral symmetry is 
broken. But even if chiral symmetry breaking is not surprising,
it is not a priori clear whether the observed pattern of 
chiral symmetry breaking and confinement is required on general 
grounds, or whether it is a particular dynamical feature of QCD. 
Some obvious questions are: Are all asymptotically free gauge 
theories confining? Does confinement imply chiral symmetry breaking 
(or vice versa)? Is the symmetry breaking pattern $SU(3)_L
\times SU(3)_R\to SU(3)_V$ unique? 

 An interesting context in which these questions can be studied
is the phase diagram of QCD and supersymmetric generalizations 
of QCD as a function of $N_c$ and $N_f$, see Fig.~\ref{fig_bigpic}. 
For our purposes supersymmetric QCD is simply a QCD-like theory 
with extra fermions in the adjoint representation and extra colored 
scalar fields. Including supersymmetric theories is useful because 
supersymmetry provides additional constraints that determine the 
symmetries of the ground state. The following interesting results 
have been obtained: 
 
 1) In supersymmetric QCD there is a window $N_c+1<N_f<3N_c$ in 
which the theory is asymptotically free but not confining 
\cite{Intriligator:1995au}. There are several reasons to believe 
that such a window exists in QCD, too. One is the fact that as a 
function of the number of flavors the second coefficient of the 
beta function changes sign before the first one does \cite{Banks:1981nn}. 
In this regime the coupling constant flows to a finite value at large 
distance and the theory is scale invariant. 

 2) Supersymmetric QCD also provides examples for theories that have 
confinement but no chiral symmetry breaking. This happens for $N_f=
N_c+1$. This theory contains both massless mesons and massless baryons. 
An important constraint is provided by the 't Hooft anomaly matching 
conditions \cite{tHooft:1980xb,Peskin:1982}. In QCD these relations 
show that confinement without chiral symmetry breaking is a possibility 
for $N_f=2$, but ruled out for $N_f>2$. 

 3) The 't Hooft consistency conditions also provide constraints
on the symmetry breaking pattern. In QCD these conditions are 
not sufficiently strong to fix the ground state completely, but 
one can show that $SU(3)_L\times SU(3)_R \to SU(3)_V$ is favored
in the limit $N_c\to\infty$ \cite{Coleman:1980mx}.

 4) One can show that in QCD chiral symmetry breaking implies
a non-zero quark condensate \cite{Kogan:1998zc}. In particular, one
can rule out the possibility that $\langle \bar{\psi}\psi\rangle
=0$, but $\langle (\bar{\psi}\psi)^2\rangle\neq 0$.

\section{QCD at finite Temperature}
\label{sec_T}
\subsection{General Arguments}
\label{sec_arg}

 In this section we shall discuss the phase structure of QCD
at non-zero temperature. We begin by reviewing some general 
arguments in favor of the existence of a critical temperature 
$T_c$ above which quarks and gluons are deconfined and chiral 
symmetry is restored. 

 Asymptotic freedom clearly suggests that the high temperature phase 
is a weakly interacting plasma \cite{Collins:1974ky,Shuryak:1977ut}. 
Consider a non-interacting gas of quarks and gluons at high temperature. 
The typical momenta are on the order of the temperature, $p\sim 3T$, 
and the density is $n\sim T^3$. Now imagine that we turn on the 
coupling. Does this lead to a qualitative change in the system? 
Quarks and gluon can scatter but since the typical momenta are large 
a significant change in the momentum of the scattered particles 
requires a large momentum transfer. Asymptotic freedom implies 
that the effective coupling at this scale is small, and that 
large angle scattering events are rare. If the change of momentum 
is small then the scattering involves large distances and the interaction 
is modified by the dense medium. We will see below that the quark-gluon 
medium screens the interaction and that the effective interaction
is again weak. There is a small subtlety here, as static magnetic
interactions are not screened. This implies that high temperature
QCD has a genuinely non-perturbative sector, but this sector is 
not important as far as bulk properties of the high temperature 
phase are concerned. We conclude that the assumption of a weakly 
interacting quark-gluon system at high temperature leads to a
self consistent picture. Since this system will exhibit medium 
effects, such as damping and screening, collective modes, 
etc.~that are typical of plasmas it was termed the quark gluon 
plasma (QGP) \cite{Shuryak:1977ut,Shuryak:1978ij}. 

 It is instructive to consider a simple model of the equation 
of state. The pressure and energy density of non-interacting 
massless particles is 
\be
P = \epsilon/3, \hspace{0.5cm}
\epsilon = g\frac{\pi^2}{30} T^4 
\left\{ \begin{array}{cl}
 1 & {\rm bosons} \\
7/8 & {\rm fermions } \end{array}\right. ,
\ee
where $g$ is the number of degrees of freedoms. In a quark 
gluon plasma we have $g_q=4N_fN_c$ quarks and $g_g=2(N_c^2-1)$
gluon degrees of freedom. For $N_f=2$ we get $g_{eff}=g_g+
7g_q/8=37$ and
\be
 P =\frac{37\pi^2}{90}T^4.
\ee
At low temperature the relevant degrees are Goldstone bosons. Near 
$T_c$ we can assume that Goldstone bosons are approximately massless 
and the number of degrees of freedom is $g=(N_f^2-1)$. For $N_f=2$
we get $g=3$ and
\be
 P =\frac{3\pi^2}{90}T^4.
\ee
This result seems to show that the pressure in the low temperature
phase is always smaller than the pressure in the high temperature
phase. This cannot be right as it would imply that the phase with
chiral symmetry breaking is never favored. The problem is that 
there are non-perturbative effects in the low temperature phase.
These effects give rise to a negative vacuum energy and a positive 
vacuum pressure. Lorentz invariance implies that the vacuum 
energy momentum tensor is of the form $T_{\mu\nu}=Bg_{\mu\nu}$
and
\be
\label{bag_const}
\epsilon_{vac} = -P_{vac} = +B .
\ee
In QCD the vacuum energy satisfies the trace anomaly relation
\be
\label{trace_an}
\epsilon_{vac} = -\frac{b}{32}\langle \frac{\alpha}{\pi} G^2\rangle 
  \simeq -0.5 \ {\rm GeV}/{\rm fm}^3.
\ee
The numerical value comes from QCD sum rule determinations of 
the gluon condensate \cite{Shifman:1978bx} $\langle \alpha G^2\rangle$ 
and has a considerable uncertainty. We can now obtain an estimate of 
the transition temperature by requiring that the pressure in the 
low temperature phase equals the pressure in the quark gluon 
phase. We find
\be
\label{tc_bag}
T_c = \left(\frac{45B}{17\pi^2}\right)^{1/4} \simeq 180\ {\rm MeV}
\ee
We can also determine the critical energy density. The energy 
densities just below and just above the critical temperature are
given by
\bea
\epsilon(T_c^-)&=&\frac{3\pi^2}{30}T_c^4\simeq 130 \ {\rm MeV}/{\rm fm}^3 , \\
\epsilon(T_c^+)&=&\frac{37\pi^2}{30}T_c^4+B\simeq 2000 \ {\rm MeV}/{\rm fm}^3.
\eea
We observe that the energy density in the QGP exceeds 1 ${\rm GeV}/{\rm fm}^3$.
This should be compared to the energy density in cold nuclear matter which 
is about 150 ${\rm MeV}/{\rm fm}^3$.

 An independent estimate of the transition temperature can be obtained
using the chiral effective theory. We saw that the chiral condensate
is related to the mass dependence of the vacuum energy. At tree level
and to leading order in the quark masses the condensate is given by 
the coefficient $B$ in the chiral lagrangian. We can also calculate 
corrections to this result due to thermal Goldstone boson fluctuations.
At leading order it is sufficient to consider the free energy of a
non-interacting pion gas
\be
\label{z_pi}
F = (N_f^2-1)T\int \frac{d^3p}{(2\pi)^3} \log
  \left( 1- e^{-E_\pi/T} \right), 
\ee
where $E_\pi=\sqrt{p^2+m_\pi^2}$. The quark condensate is 
$\langle\bar\psi\psi\rangle = (N_f)^{-1}\partial F/\partial m$. 
Equation (\ref{z_pi}) depends on the quark mass only through the 
pion mass. Using the Gell-Mann-Oakes-Renner relation (\ref{GMOR})
we find \cite{Gasser:1986vb}
\be
\langle\bar\psi\psi\rangle_T = 
\langle\bar\psi\psi\rangle_0 \left\{ 1-\frac{N_f^2-1}{3N_f}
\left(\frac{T^2}{4f_\pi^2}\right)+\ldots\right\}.
\ee
This result indicates that the chiral condensate vanishes at 
a critical temperature 
\be 
\label{tc_chi}
T_c\simeq 2f_\pi\sqrt{\frac{3N_f}{N_f^2-1}}\simeq 200 \, 
{\rm MeV} \;\; (N_f=3),
\ee
which is roughly consistent with the estimate obtained in 
equ.~(\ref{tc_bag}).

\subsection{Chiral Symmetry Restoration}
\label{sec_csr}

 In the vicinity of the phase transition QCD is genuinely 
non-perturbative and it is hard to improve on the rough 
estimates provided in the previous section. One possibility
for determining the transition temperature and elucidating 
the nature of the phase transition is the use of large scale 
numerical simulations. We will discuss this approach in 
Sect.~\ref{sec_lqcd}. Before we do so we would like to review
certain general statements about the phase transition that 
follow from the symmetries of the low and high temperature 
phase. 

\begin{table}[t]
\tbl{Correspondence between the chiral phase transition in QCD 
and the ferromagnetic transition in a four-component magnet.}
{\begin{tabular}{clccl}
$SU(2)_L\times SU(2)_R$ & QCD  & \hspace{0.5cm} & 
$O(4)$ &  magnet \\
$\langle\bar\psi\psi\rangle$  & $\chi$ condensate  & &
$\vec{M} $  & magnetization  \\
$m_q$  & quark mass  & & 
$H_3$  & magnetic field  \\
$\vec\pi$  & pions  & & 
$\vec\phi$  & spin waves  
\end{tabular}
\label{tab_uni}}
\end{table}

 We begin with the chiral phase transition. We shall assume
that the chiral transition is a second order phase transition,
i.e.~the order parameter goes to zero continuously. We will 
explore the consequences of this assumption and check whether
it leads to a consistent picture. In the vicinity of a second
order phase transition the order parameter fluctuates on all
length scales and the correlation length diverges. This means
that the details of the interaction are not important, and 
only the symmetry of the order parameter matters.

 In QCD with two flavors the order parameter is a $2\times 2$
matrix $U^{fg}=\langle \bar{\psi}^f_L\psi^g_R\rangle$. We can 
define a four component vector $\phi^a$ by $U^{fg}=\phi^a
(\tau^a)^{fg}$ with $\tau^a=(\vec{\tau},1)$. Chiral transformations 
$(L,R)\in SU(3)_L\times SU(3)_R$ correspond to rotations $\phi^a
\to R^{ab}\phi^b$ with $R\in SO(4)$. In the low temperature phase
chiral symmetry is broken and $\langle\phi^a\rangle = \sigma 
\delta^{a0}$. Near $T_c$ the order parameter is small and we
we can expand the free energy in powers of the order parameter 
and its derivatives. Chiral symmetry implies that only the length 
of $\phi^a$ can enter. To order $\phi^4$ and to leading order 
in gradients of the fields we have 
\be
\label{f_lg}
F = \int d^3x\, \left\{ \frac{1}{2}(\vec\nabla\phi^a)^2
  +\frac{\mu^2}{2}(\phi^a\phi^a) +\frac{\lambda}{4}
  (\phi^a\phi^a)^2 + \ldots \right\},
\ee
where $\mu^2,\lambda$ are parameters that depend on the temperature. 
Equ.~(\ref{f_lg}) is the Landau-Ginzburg effective action. Note 
that fluctuations of the order parameter are dominated by static
fields $\phi^a(\vec{x},t)=\phi^a(\vec{x})$. This will be explained
in more detail in Sect.~\ref{sec_pqcd}. The main observation is 
that fluctuations with energy much smaller than $\pi T$ are 
described by a three dimensional theory. 

  Stability requires that $\lambda>0$. Below the critical temperature 
$\mu^2<0$ and chiral symmetry is broken. At $T_c$ the parameter $\mu^2$ 
changes sign and we can write $\mu^2=\mu_0^2 t$ where $t=(T-T_c)/T_c$
is the reduced temperature. As a first approximation we can 
ignore fluctuations and study the Landau-Ginzburg action in the 
mean field approximation. In that case the chiral order parameter 
goes to zero as $\langle\phi^0\rangle \sim t^{1/2}$. This 
result is modified if fluctuations are included. This can 
be done using renormalization group methods or numerical
simulations. These methods also demonstrate that near $T_c$
higher order operators not included in equ.~(\ref{f_lg})
are indeed irrelevant. The results are 
\be
\begin{array}{rclrcl}
C &\sim & t^{-\alpha} &  \hspace{0.1\hsize}\alpha &=& -0.19, \\
\langle\bar\psi\psi\rangle 
  &\sim & t^\beta     &                    \beta  &=& 0.38, \\
m_\pi &\sim & t^\nu   &                    \nu    &=& 0.73,
\end{array}
\ee
where $C$ is the specific heat and the Goldstone boson mass
$m_\pi$ is defined as the inverse correlation length of spatial 
fluctuations of the pion field. 

 The coefficients $\alpha,\beta,\nu$ are called the critical 
indices of the phase transition. These coefficients are 
universal, i.e.~independent of the details of the microscopic
details. For example, the critical indices of the chiral 
phase transition in QCD agree with the critical indices
of a four-component magnet in $d=3$ space dimensions, see
Table \ref{tab_uni}.

 In QCD with $N_f=3$ flavors the order parameter is a $3\times 3$
matrix $U^{fg}$. The main new ingredient in the Landau-Ginzburg
theory is a chirally invariant cubic term $\det(U)+{\rm h.c}$.
It is easy to check that the cubic term will lead to an effective 
potential that has two degenerate minima at the transition 
temperature. This implies that the transition is first order
and the order parameter goes to zero discontinuously. Even 
if the coefficient of the cubic term is zero initially, 
fluctuations tend to generate a cubic interaction as $T\to T_c$.
A transition of this type is called fluctuation induced 
first order. 

\subsection{Deconfinement}
\label{sec_dec}

  It is not immediately obvious how to identify the symmetry 
associated with the deconfinement transition. In order to make 
contact with the lattice formulation discussed in Sect.~\ref{sec_lqcd} 
we will study this question in euclidean space, i.~e.~with the 
time variable analytically continued to imaginary time, $t\to it$.
We first derive an expression for the potential between two
heavy quarks. Consider the Dirac equation for a massive quark 
in a gluon background field
\be
\left(\partial_0-igA_0 -\vec\alpha(i\vec\nabla+g\vec{A})
+\gamma_0 M\right)\psi = 0 .
\ee
In the limit $m_Q\to\infty$ we can ignore the spatial components
of the covariant derivative. The quark propagator is 
\be
S(x,x')\simeq \exp\left(ig\int A_0dt\right) \left(\frac{1+\gamma_0}{2}\right)
e^{-m(t-t')}\delta(\vec{x}-\vec{x}').
\ee
The heavy quark potential is related to the amplitude for creating 
a heavy quark-anti-quark pair at time $t=0$, separating the pair
by a distance $R$, and finally annihilating the two quarks at time 
$t=T$. The amplitude is related to the Wilson loop
\be
\label{w_loop}
 W(R,T)=\exp\left(ig\oint A_\mu dz_\mu\right),
\ee
where the integration contour is a $R\times T$ rectangle and
we have dropped the Dirac projection operators $(1+\gamma_0)/2$.
If $T\gg R$ the amplitude is dominated by the ground state and we 
expect $W(R,T)\sim \exp(-V(R)T)$ where $V(R)$ is the heavy quark 
potential. Confinement implies that $V(R)\sim kR$ and the Wilson 
loop satisfies an area law
\be
\label{area_law}
 W(R,T)\sim \exp(-k A), . 
\ee
where $A=RT$. In order to construct a local order parameter 
we consider the  Polyakov line
\be
\label{p_line}
 P(\vec{x})=\frac{1}{N_c}{\rm Tr}[L(\vec{x})] = 
  \frac{1}{N_c}P{\rm Tr}
  \left[\exp\left(ig\int_0^\beta A_0 dt\right)\right].
\ee
As we will explain in more detail in the next section 
gauge fields are periodic in imaginary time with period 
$\beta=1/T$. The Polyakov line can be interpreted as the
free energy of a single quark, $\langle P\rangle \sim 
\exp(-m_Q \beta)$. In the confined phase we expect that the 
energy of an isolated quark is infinite, while in the 
deconfined phase it is finite. This implies that 
\be
\langle P\rangle = 0 \hspace{0.5cm}{\rm confined},\hspace{1.5cm}
\langle P\rangle \neq 0 \hspace{0.5cm}{\rm deconfined}.
\ee
The global symmetry of the order parameter is \cite{Gross:1980br} 
$P\to zP$ with $z=\exp(2\pi ki/N_c)\in Z_{N_c}$. Since 
$Z_{N_c}$ is the center of the gauge group $SU(N_c)$ this 
is sometimes called the center symmetry. Color singlet 
correlation functions always involve combinations of
Polyakov lines that are invariant under center transformations.
A heavy baryon correlation function, for example, is of the 
form ${\rm tr}[(L(\vec{x}))^{N_c}]$ and is invariant 
because $z^{N_c}=1$. A non-zero expectation value for the 
Polyakov line, on the other hand, breaks center symmetry.  

\begin{figure}[t]
\bc\includegraphics[width=8.0cm]{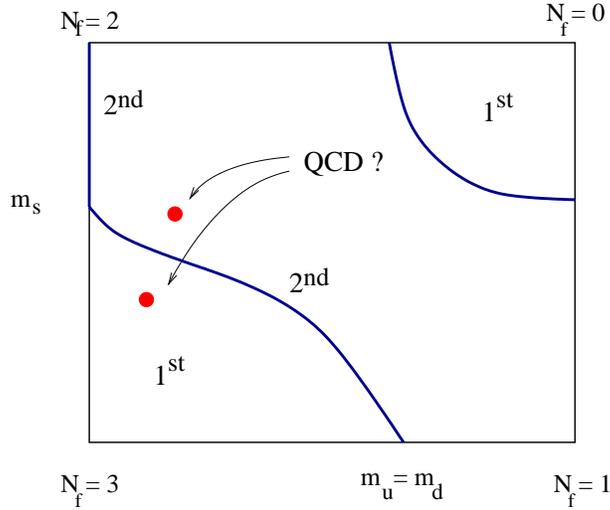}\ec
\caption{\label{fig_uni}
Phase diagram of QCD in the $m-m_s$ mass plane. The plot
shows the universality class of the chiral/deconfinement 
transition for different values of $m,m_s$.}
\end{figure}

 Note that the symmetry is broken in the high temperature 
phase and restored in the low temperature phase. This might 
seem somewhat unusual, but there are examples of spin 
systems that have an equivalent ``dual'' description in 
terms of a gauge theory \cite{Creutz:1984mg}. In the dual theory 
the high and low temperature phases are interchanged. 
 
 The $Z_{N_c}$ Landau-Ginzburg action is given by
\cite{Svetitsky:1982gs}
\be
\label{f_pol}
F = \int d^3x\, \left\{ \frac{1}{2}|\vec\nabla P|^2
  +\mu^2|P|^2  + g{\rm Re}(P^3) +\lambda |P|^4 + \ldots \right\}.
\ee
The cubic term is allowed only if $N_c=3$. As in the case
of the chiral phase transition we expect the cubic term 
to drive the transition first order. The two color theory 
is in the equivalence class of the $Z_2$ Ising model, which 
is known to have a second order transition. The three color 
theory is in the equivalence class of a three state  Potts
model, which does indeed show a first order transition.

 The phase structure as a function of the light quark mass
$m=m_u=m_d$ and the strange quark mass $m_s$ is summarized
in Fig.~\ref{fig_uni}. The lower left hand corner of the 
diagram is $m=m_s=0$ and corresponds to three massless 
quarks. In that case we expect a first order chiral phase
transition. Along the diagonal we have $m=m_s$ and the 
$SU(3)_V$ flavor symmetry is exact. As the quark masses
increase the strength of the first order transition becomes
weaker and the transition eventually ends at a second 
order critical point. If the light quarks are kept massless
as the strange quark mass is increased the endpoint of 
the first order transition is a tricritical point at 
which the the endpoint of the first order transition in 
the three flavor theory meets the second order transition 
of the two flavor theory. This transition turns into a
smooth crossover as soon as the light quarks are given 
a small mass. 

 The upper right hand corner of the plot is $m=m_s\to\infty$
and corresponds to the pure glue theory which has a first 
order transition. Dynamical quarks break the $Z_{N_c}$ 
symmetry and the strength of the first order transition
decreases as the quark masses are lowered. The first order
transition in the pure glue theory ends on a line of 
second order transitions. 

 We do not know with certainty where the physical point is 
located on this phase diagram. Lattice calculations currently 
favor the possibility that the phase transition is in the 
crossover region, closer to the first order chiral transition
than to the first order deconfinement transition. 

\begin{figure}[t]
\bc\includegraphics[width=5.0cm]{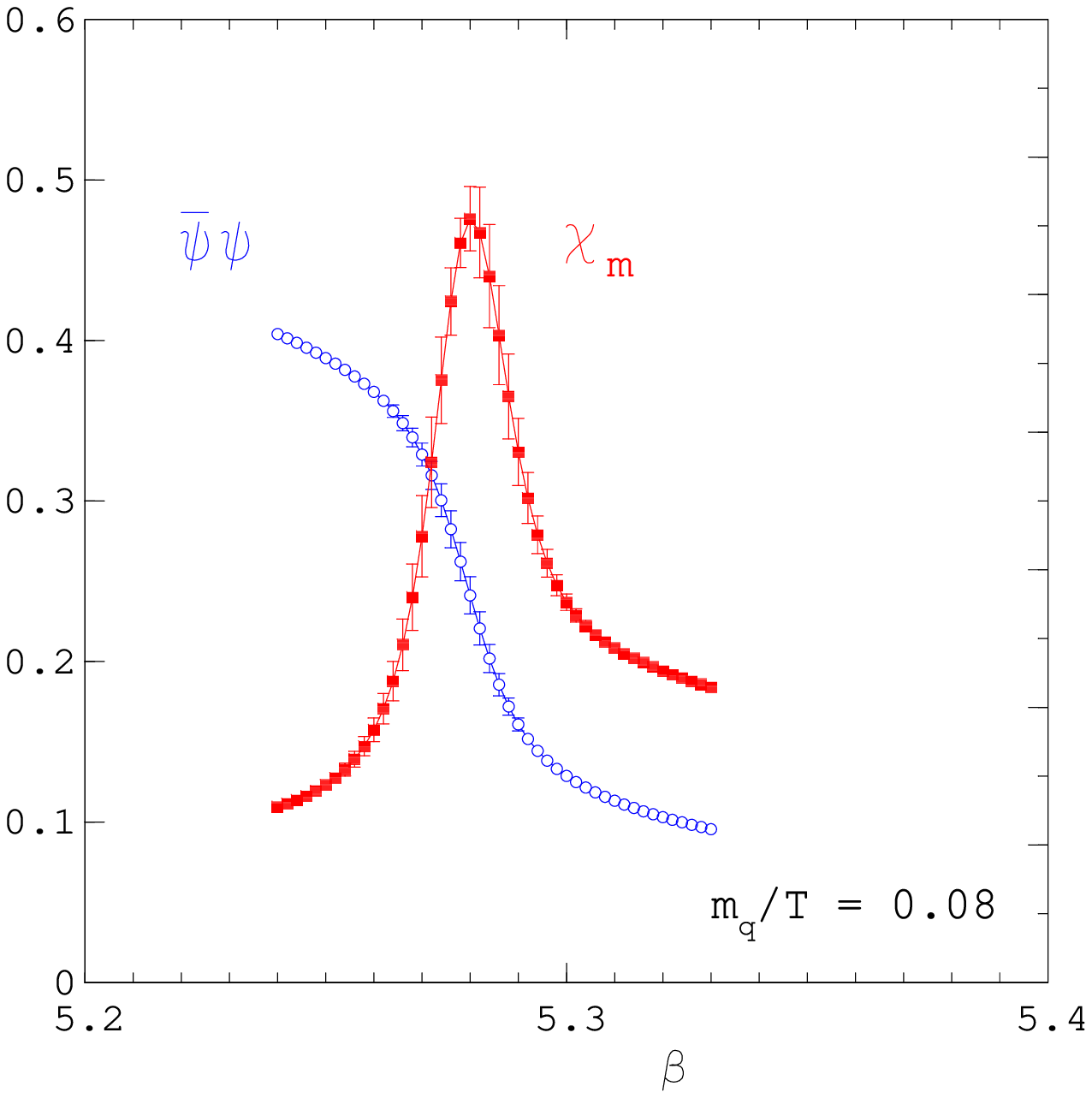}
\includegraphics[width=5.0cm]{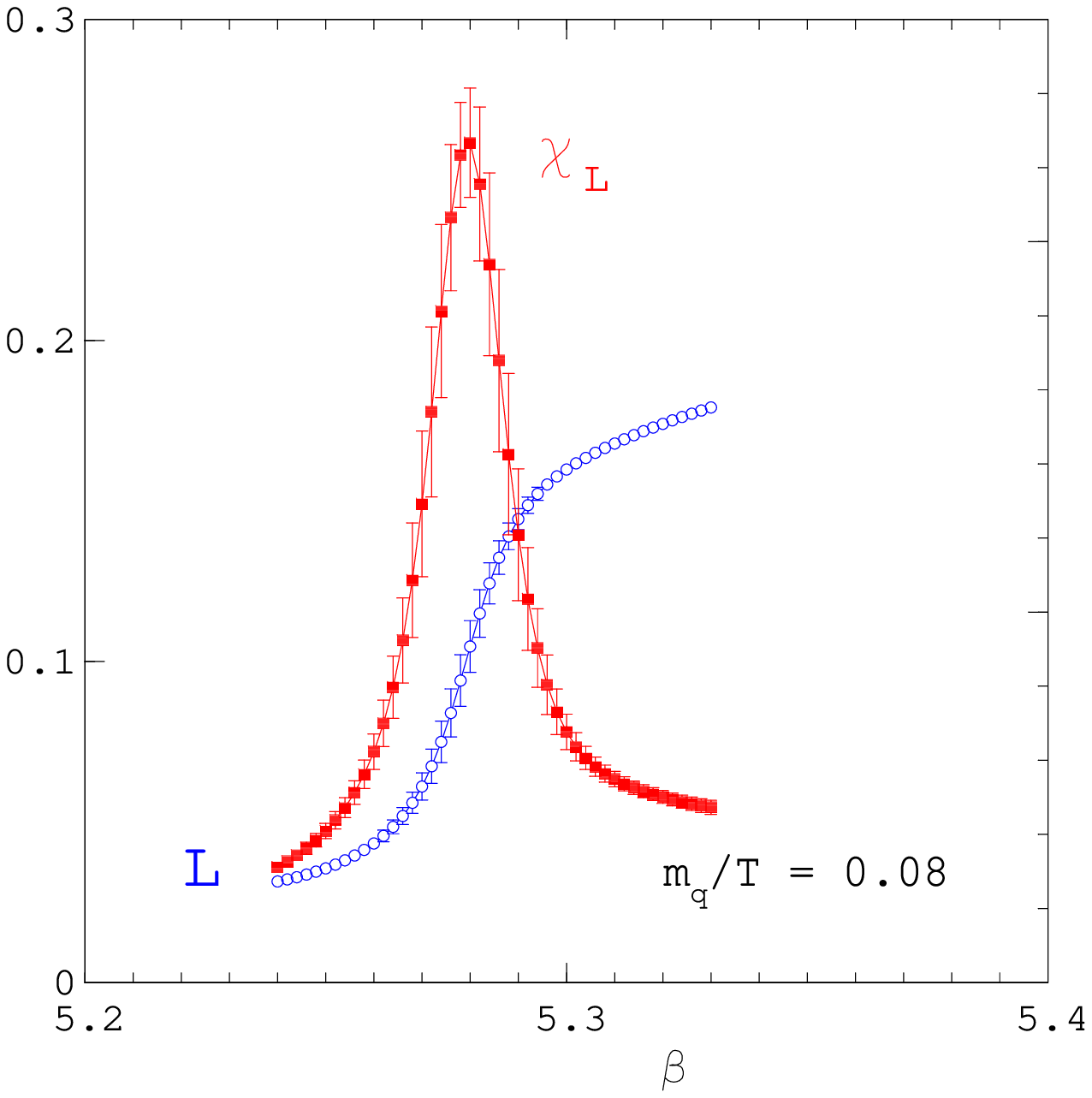}\ec
\caption{\label{fig_sus}
Chiral and deconfinement transitions for $N_f=2$ dynamical
quark flavors, from (Karsch 2002). The two figure show the chiral 
condensate and the Polyakov line as a function of the bare coupling 
$\beta=6/g^2$ (In this figure, $\beta$ is not $1/T$). Asymptotic 
freedom implies that on a fixed lattice $N_\tau\times N_\sigma^3$ 
increasing $\beta$ corresponds to increasing the temperature. The 
susceptibilities $\chi_m,\chi_L$ are related to order parameter
fluctuations, e.g. $\chi_L=n_\sigma^3(\langle L^2\rangle -
\langle L\rangle^2)$. }
\end{figure}

 We should emphasize that Fig.~\ref{fig_uni} focuses on regions
of the phase diagram in which there is a sharp phase transition
and the order parameter is a non-analytic function at $T_c$. The
figure should not be taken to imply that the chiral and 
deconfinement transitions are completely separate phenomena, 
or that the transition cannot be observed in the crossover 
region. Fig.~\ref{fig_sus} shows the order parameter as 
well as the order parameter susceptibilities for the chiral
and deconfinement transitions. The results were obtained 
from lattice calculations with semi-realistic values of the 
quark masses. We observe that even though both transitions 
are crossovers clear peaks in the susceptibilities are still 
visible. We also note that the chiral and deconfinement 
transitions occur at the same temperature. This can be 
understood in models that include a non-zero coupling between 
the effective actions for the chiral and deconfinement order 
parameters \cite{Digal:2000ar,Mocsy:2003qw,Ratti:2005jh}.

\subsection{Lattice QCD}
\label{sec_lqcd}

 Symmetry arguments cannot determine the critical temperature, 
the critical energy density, and many other properties of 
matter near the phase transition. In order to compute these
properties we have to rely on numerical simulations of the 
QCD partition function 
\be
Z = {\rm Tr}[e^{-\beta H}], \;\; \beta=1/T
\hspace{0.5cm}F=TV{-1}\log(Z) ,
\ee
where $H$ is the QCD Hamiltonian, $\beta$ is the inverse temperature
and $F$ is the free energy. We can write the partition function 
as a quantum mechanical evolution operator $Z = {\rm Tr}[e^{-i
(-i\beta) H}]$ with imaginary time $\tau=-i\beta$. The evolution
operator can be written as a path integral 
\be
\label{z_path}
Z =\int dA_\mu d\psi d\bar{\psi}
   \exp\left(-\int_0^\beta d\tau \int d^3x\ {\mathcal L}_E\right),
\ee
where ${\mathcal L_E}$ is the imaginary time (euclidean) lagrangian 
and we have to impose (anti)periodic boundary conditions on the 
quark and gluon fields
\be
 A_\mu(\vec{x},\beta)=A_\mu(\vec{x},0), \hspace{0.3cm}
 \psi(\vec{x},\beta)=-\psi(\vec{x},0).
\ee
The path integral is an infinite dimensional integral. In order 
to perform numerical simulations we have to discretize space 
and time and introduce a $N_\tau\times N_\sigma^3$ lattice with 
lattice spacing $a$. In order to maintain exact gauge invariance
on a discrete lattice the gauge fields are discretized in terms
of the link variables
\be
U_\mu(n)=\exp(igaA_\mu(n)),
\ee
where $n=(n_\tau,n_x,n_y,n_z)$ labels lattice sites and $\mu=1,
\ldots,4$. In terms of the link variables it is easy to define a 
gauge covariant discrete derivative
\be
D_\mu\phi \to \frac{1}{a}[U_{-\mu}(n)\phi(n+\mu)-\phi(n)],
\ee
where $n+\mu$ is the lattice site reached by starting at $n$
and doing one hop in the $\mu$ direction. $\phi(n)$ is a scalar 
field in the fundamental representation. The action of the pure 
gauge theory is given by
\be
\label{wilson}
 S =  \frac{2}{g^2}\sum_{n,\mu} {\rm Re\ Tr} 
  \left[1- U_\mu(n)U_\nu(n+\mu) U_{-\mu}(n+\mu+\nu)U_{-\nu}(n+\nu) 
     \right], 
\ee
which involves a loop of link variables called a plaquette. It is
easy to check that equ.~(\ref{wilson}) reduces to the continuum
action as $a\to 0$. Fermion fields $\psi(n)$ are discretized
on lattice sites. There are some subtleties in discretizing chiral 
fermions which we do not discuss here \cite{Chandrasekharan:2004cn}. 
The action is bilinear in the fermion fields. This means that the 
integral over the fermion fields can be done exactly and we are left 
with the determinant of a matrix that depends only on the link 
variables. The lattice representation of the partition function 
is 
\be 
 Z = \int\prod_{n,\mu}dU_\mu(n) \det (M(U)) e^{-S},
\ee
where $M(U)$ is the fermion matrix. The partition function depends
on the number of lattice sites $N_\tau,N_\sigma$, the bare 
coupling constant $g$, and the dimensionless quark masses $ma$.

\begin{figure}[t]
\bc\includegraphics[width=10.0cm]{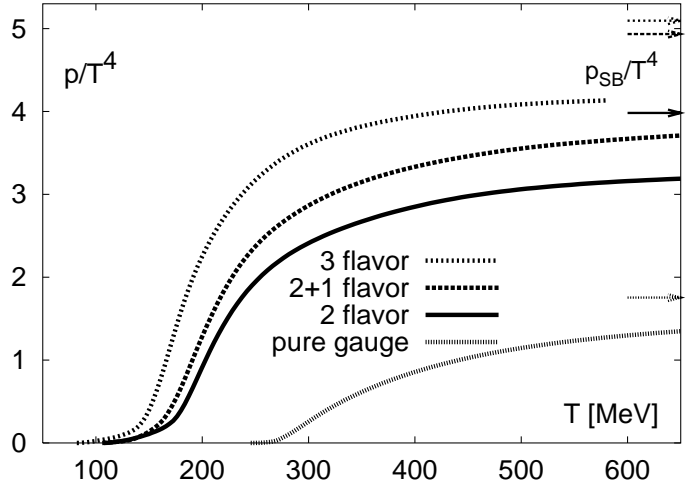}\ec
\caption{\label{fig_eos}
Equation of state obtained in lattice calculations with
$N_f=0,2,2+1,3$ flavors, from Karsch (2002). The 2+1 curve 
refers to two light flavors and one intermediate mass flavor. 
The pressure is given in units of $T^4$. The arrows indicate 
the Stefan-Boltzmann limits.  }
\end{figure}

 Note that the partition function has no explicit dependence
on the lattice spacing. Asymptotic freedom implies that the 
bare coupling should go to zero as $a\to 0$ and the continuum 
limit corresponds to $g\to 0$. Asymptotically
\be 
\label{lam_lat}
 a\Lambda_{lat}=\exp(-8\pi^2/(bg^2)) ,
\ee
where $b$ is the first coefficient of the beta function, see
equ.~(\ref{as_fr}), and $\Lambda_{lat}$ is the QCD scale 
parameter on the lattice. $\Lambda_{lat}$ can be related 
to the continuum scale parameter by a perturbative calculation
\cite{Creutz:1984mg}. In practice the lattice spacing is not small 
to enough to use the perturbative result equ.~(\ref{lam_lat}) 
and $a$ is determined by from a physical quantity like 
the string tension or the rho meson mass. Once the lattice
spacing is known the temperature is determined by $T=1/
(N_\tau a)$.

 Lattice results for the order parameter and the equation
of state are shown in Figs.~(\ref{fig_sus},\ref{fig_eos}).
Current results for the transition temperature are 
\cite{Laermann:2003cv}
\be 
T_c(N_f\!=\!2)= (173\pm 8)\ {\rm MeV},\hspace{0.5cm}
T_c(N_f\!=\!0)= (271\pm 2)\ {\rm MeV}, 
\ee
where the errors are purely statistical. The equation of 
state shows a rapid rise in the pressure near $T_c$, but 
the pressure remains significantly below the free gas 
limit even at $T\sim (2-3)T_c$. 

\subsection{Perturbative QCD}
\label{sec_pqcd}

\begin{figure}[t]
\bc\begin{minipage}{5.5cm}
\includegraphics[width=4.7cm]{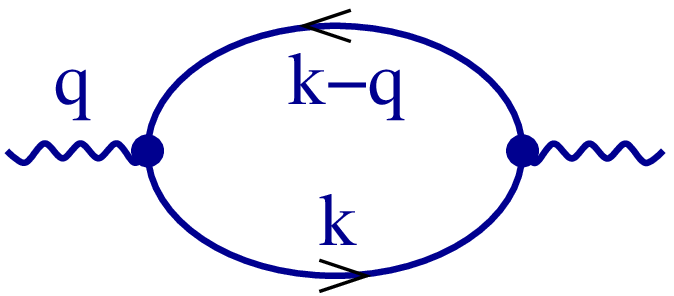}
\vspace*{0.7cm}
\end{minipage}\begin{minipage}{5.5cm}
\includegraphics[width=5.0cm]{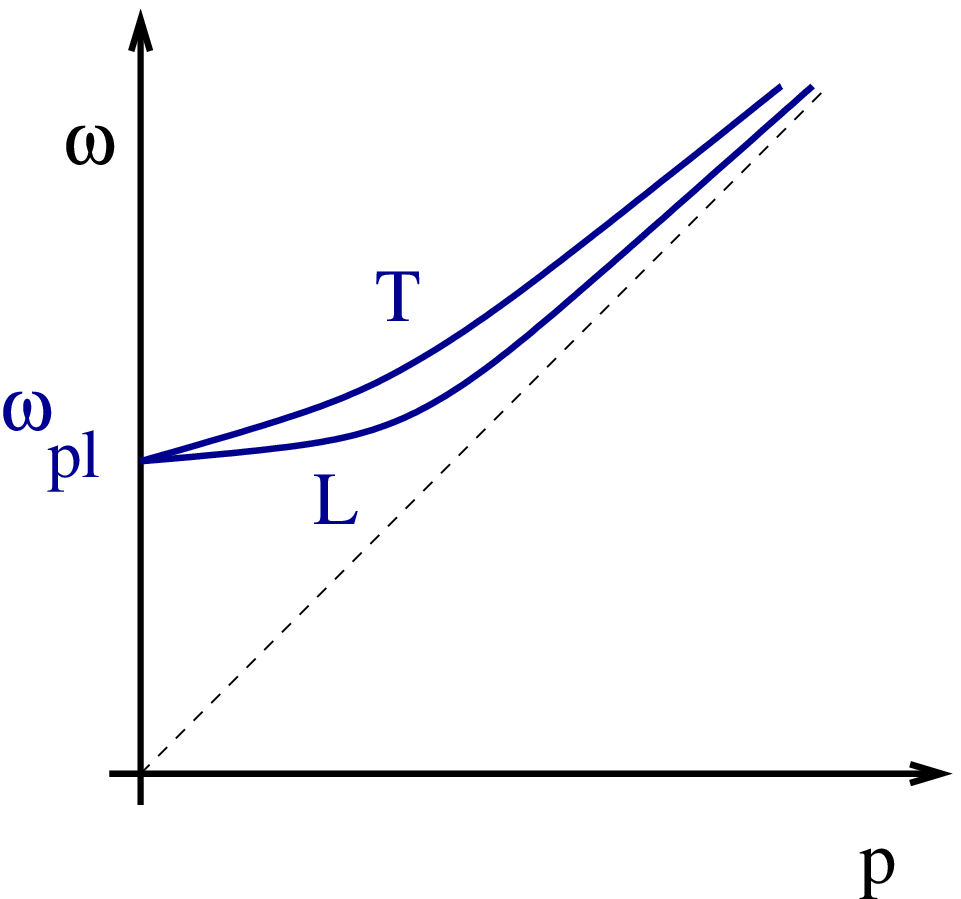}
\end{minipage}\ec
\caption{\label{fig_pol}
One loop contribution to the photon polarization tensor 
(left panel) and plasmon dispersion relation in a hot QED 
plasma. }
\end{figure}

 At temperatures significantly above $T_c$ quarks and gluons 
are weakly coupled and perturbative methods are expected 
to be useful. The starting point of the perturbative 
expansion at non-zero temperature is the path integral 
representation of the QCD partition function given in 
equ.~(\ref{z_path}). The only difference as compared to
the zero temperature result is the fact that the fields
satisfy periodic boundary conditions in imaginary time. 
Consider the Fourier representation of the gluon field
\be 
A_\mu(\vec{x},\tau) = \sum_n\int d^3k\;  A_\mu^n(\vec{k})
 e^{-i(\vec{k}\vec{x}+\omega_n\tau)}.
\ee
We can also write down an analogous expansion for fermions. 
The boundary conditions imply that the allowed frequencies, 
called the Matsubara frequencies, are discrete. We have
\be
\begin{array}{rcll}
\omega_n &=& 2\pi n T                    & {\rm bosons} \\
\omega_n &=& (2 n+1)\pi T\hspace{0.2cm}  & {\rm fermions}
\end{array}.
\ee
The only change in the Feynman rules is that continuous
energy variables are replaced by discrete Matsubara frequencies,
$p_0\to i\omega_n$, and that integrals over energy are replaced 
by discrete sums 
\be
\int\frac{d^4p}{(2\pi)^4}\to
T\sum_n \int \frac{d^3p}{(2\pi)^3}.
\ee
Typical sums that appear in one-loop calculations in the 
Matsubara formalism are 
\bea
\label{mats_sum1}
\sum _k \frac{1}{x^2+k^2} &=& 
\frac{2\pi}{x}\left( \frac{1}{2}  +\frac{1}{e^{2\pi x}-1}\right), \\
\label{mats_sum2}
\sum _k \frac{1}{x^2+(2k+1)^2} &=& 
\frac{\pi}{x}\left( \frac{1}{2}  -\frac{1}{e^{\pi x}+1}\right) .
\eea
We observe that performing sums over Matsubara frequencies leads
to Bose-Einstein and Fermi-Dirac distribution functions. 

 As an application of the finite temperature formalism we wish 
to study the one-loop correction to the gluon propagator in a hot 
QCD medium. For simplicity we begin with the analogous problem 
in a hot QED plasma. The photon polarization function is (see
Fig.~\ref{fig_pol})
\be
\label{ph_pol}
\Pi_{\mu\nu}(q) = e^2T\sum_n\int \frac{d^3k}{(2\pi)^3}
  {\rm tr}[\gamma_\mu k\slash\gamma_\nu(k\sslash-q\sslash ) ]
   \Delta(k)\Delta(k-q),
\ee
where $\Delta(k)=\omega_n^2+\vec{k}^2$. Using identities like 
equ.~(\ref{mats_sum2}) we can decompose the integral into a 
$T=0$ and a finite temperature part. In the following we will 
only consider the $T\neq 0$ terms. Thermal corrections to the 
photon propagator become important when the photon momentum is 
much smaller than the temperature. In this case we can assume 
that the loop momenta are on the order of $T$. This is called 
the hard thermal loop (HTL) approximation \cite{Braaten:1989mz}.
The photon polarization function in the HTL approximation is 
\be
\label{pi_htl}
\Pi_{\mu\nu} = 2m^2 \int\frac{d\Omega}{4\pi}
 \Big(\frac{i\omega\hat{K}_\mu\hat{K}_\nu}{q\cdot\hat{K}}
  +\delta_{\mu 0}\delta_{\nu 0}\Big) , 
\ee
where $m^2=e^2T^2/6$, $\hat{K}=(-i,\hat{k})$ and $d\Omega$ is an 
integral over the direction of $\hat{k}$. In the case of the 
gluon propagator in QCD there are extra graphs generated by the
three and four-gluon vertices as well as ghost contributions but, 
remarkably, the structure of the HTL polarization function is 
unchanged. In QCD the parameter $m^2$ is given by $m^2=g^2T^2
(1+N_f/6)$.

\begin{figure}[t]
\bc\includegraphics[width=11.0cm]{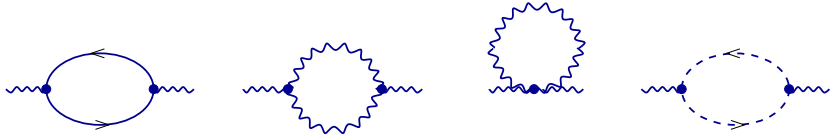}\ec
\caption{\label{fig_htl}
One loop contribution to the gluon polarization tensor 
in the quark gluon plasma. Solid lines are quark propagators, 
wavy lines are gluons, and dashed lines are ghosts.}
\end{figure}

 Insertions of the polarization function into the photon (gluon) 
propagator form a simple geometric series. The resummed photon 
propagator is 
\be 
\label{d_res}
D_{\mu\nu}=  \frac{1}{(D_{\mu\nu}^0)^{-1}+\Pi_{\mu\nu}}.
\ee
This result can be used to study the interaction between two 
charges in the plasma. The Coulomb interaction is determined
by the Fourier transform of the static propagator
\be
\label{v_scr}
V(r)=e\int \frac{d^3q}{(2\pi)^3}\frac{e^{iqr}}{\vec{q}^{\,2}+\Pi_{00}}
 \simeq -\frac{e}{r}\exp(-m_D r) ,
\ee
where $m_D^2=2m^2$ is the Debye mass and we have used $\Pi_{00}(0,
\vec{q}\to 0)=m_D^2$. Equ.~(\ref{v_scr}) shows that the Coulomb 
interaction is screened at distances $r\sim m_D^{-1} \sim (eT)^{-1}$. 
The mechanism for charge screening is easy to understand. A test
charge polarizes the electron-positron plasma, and the polarization
cloud screens the charge. Note that the typical distance between
charges is $r\sim T^{-1}\ll r_D$ and Debye screening is a collective
effect that involves many charge carriers. 

 The magnetic interaction is not screened, $\Pi_{ii}(0,\vec{q}\to 0)=0$.
However, if the photon energy is finite the polarization develops
an imaginary part
\be
{\rm Im}\Pi_{ii}(\omega,q) \sim \frac{\omega}{q}m_D^2\Theta(q-\omega), 
\ee
and non-static exchanges are damped. This phenomenon is called
Landau damping. The mechanism of Landau damping is a transfer 
of energy from the electromagnetic field to electrons and positrons 
in the plasma. The absence of magnetic screening implies that the 
static magnetic sector of the QCD plasma remains non-perturbative
even if the temperature is very high. 

 In order to study the propagation of collective modes in the 
plasma in more detail it is useful to split the polarization tensor 
into transverse and longitudinal components
\bea 
\label{proj}
 \Pi_{\mu\nu}(q)&=& \Pi^T(q)P^T_{\mu\nu} +\Pi^L(q)P^L_{\mu\nu}\\
 P_{ij}^T &=& \delta_{ij}-\hat{q}_i\hat{q}_j , \hspace{1cm}
 P_{00}^T = P_{0i}^T = 0,    \\
 P_{\mu\nu}^L &=& -g_{\mu\nu}+\frac{q_\mu q_\nu}{q^2}
   -P_{\mu\nu}^T .
\eea
We can study the propagation of photons by identifying the poles 
of the transverse and longitudinal components of the photon
propagator, see Fig.~\ref{fig_pol}. We observe that for large 
momenta $|\vec{q}|\gg m$ the dispersion relation is not strongly 
affected by the medium. In this limit we also find that the 
longitudinal mode has an exponentially small residue. As $\vec{q}
\to 0$ the energy of both longitudinal and transverse modes approach 
the plasma frequency $\omega_{pl}=\sqrt{2/3}\ m$.

\section{QCD at Small Density: Nuclear Matter}
\label{sec_ldense}
\subsection{Introduction}
\label{sec_dense_intro}

 In this section we study hadronic matter at non-zero 
baryon density. In QCD the numbers of all quark flavors 
are conserved individually. Once the weak interaction is 
taken into account only baryon number and electric charge
are conserved. Bulk matter has to be electrically neutral 
because the Coulomb energy of a charged system diverges
in the infinite volume limit. In hadronic matter neutrality
can be achieved by balancing the charge density in hadrons,
which is usually positive, by a finite density of electrons. 
 
\begin{figure}[t]
\bc\includegraphics[width=7.5cm]{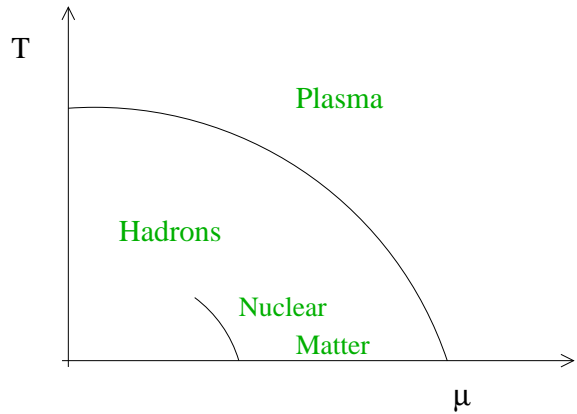}\ec
\caption{\label{fig_phase_1}
Naive phase diagram of hadronic matter as a function of the
baryon chemical potential and temperature.}
\end{figure}

 The partition function of QCD at non-zero baryon chemical 
potential is given by 
\be 
Z = \sum_i \exp\left(-\frac{E_i-\mu N_i}{T}\right),
\ee
where $i$ labels all quantum states of the system, $E_i$ and $N_i$ 
are the energy and baryon number of the state $i$. If the temperature 
and chemical potential are both zero then only the ground state 
contributes to the partition function. All other states give 
exponentially small contributions. QCD has a massgap for states 
with non-zero baryon number. This means that there is an onset
chemical potential
\be 
\mu_{\it onset}=\min_i (E_i/N_i),
\ee 
such that the partition function is independent of $\mu$ for
$\mu<\mu_{\it onset}$. For $\mu>\mu_{\it onset}$ the baryon 
density is non-zero. If the chemical potential is just above 
the onset chemical potential we can describe QCD, to first 
approximation, as a dilute gas of non-interacting nucleons. 
In this approximation $\mu_{\it onset}=m_N$. Of course, the 
interaction between nucleons is essential. Without it, we 
would not have stable nuclei. As a consequence, nuclear matter
is self-bound and the energy per baryon in the ground state 
is given by
\be 
\frac{E_N}{N}-m_N \simeq -15\,{\rm MeV}.
\ee
The onset transition is a first order transition at which 
the baryon density jumps from zero to nuclear matter saturation
density, $\rho_0\simeq 0.14\,{\rm fm}^{-3}$. The first order 
transition continues into the finite temperature plane and 
ends at a critical endpoint at $T=T_c\simeq 10$ MeV, see
Fig.~\ref{fig_phase_1}. 

 Nuclear matter is a complicated many-body system and, unlike the 
situation at zero density and finite temperature, there is little
information from numerical simulations on the lattice. This is 
related to the so-called 'sign problem'. At non-zero chemical 
potential the path integral representation of the partition 
function is 
\be
Z=\int dA_\mu \det(iD\slash +i\mu\gamma_4)e^{-S} =
\int dA_\mu e^{i\phi}|\det(iD\slash +i\mu\gamma_4)|e^{-S},
\ee
where $\phi$ is the complex phase of the fermion determinant. 
Since the determinant is complex standard Monte-Carlo techniques 
based on importance sampling fail. Recently, some progress has been 
made in simulating QCD for small $\mu$ and $T\simeq T_c$
\cite{Fodor:2001pe,deForcrand:2002ci,Allton:2002zi}, but the 
regime of small temperature remains inaccessible.

 However, if the density is very much larger than nuclear matter 
saturation density, $\rho\gg\rho_0$, we expect the problem to simplify. 
In this regime it is natural to use a system of non-interacting quarks 
as a starting point \cite{Collins:1974ky}. The low energy 
degrees of freedom are quark excitations and holes in the 
vicinity of the Fermi surface. Since the Fermi momentum is 
large, asymptotic freedom implies that the interaction between 
quasi-particles is weak. As a consequence, the naive expectation
is that chiral symmetry is restored and quarks and gluons are 
deconfined. It seems natural to assume that the quark liquid 
at high baryon density is continuously connected to the 
quark-gluon plasma at high temperature. These naive expectations
are summarized in the phase diagram shown in Fig.~\ref{fig_phase_1}.

\subsection{Fermi liquids}
\label{sec_fl}

  Before we study the high density phase in more detail we 
would like to discuss systems of nucleons at low density. For 
simplicity we will begin with pure neutron matter at densities
below nuclear matter saturation density. This problem is relevant 
to the behavior of matter near the surface of a neutron star,
which is at subnuclear densities and has a large neutron-to-proton
ratio. We will also see that pure neutron matter exhibits some
very interesting universal features which can be studied 
experimentally using trapped atomic gases. 

  If the density is low then the typical momenta are small and
neither the structure of the neutron nor the details of the
neutron-neutron interaction are important. This means that the 
system can be described by an effective lagrangian of pointlike
nucleons interacting via a short-range interaction 
\cite{Abrikosov:1963,Hammer:2000xg}. The lagrangian is 
\be 
\label{l_4f}
{\mathcal L}_0 = \psi^\dagger \left( i\partial_0 +
 \frac{\nabla^2}{2m} \right) \psi 
 - \frac{C_0}{2} \left(\psi^\dagger \psi\right)^2 .
\ee
The coupling constant $C_0$ is related to the scattering 
length, $C_0=4\pi a/m$. Note that $C_0>0$ corresponds to 
a repulsive interaction, and $C_0<0$ is an attractive interaction.
The lagrangian equ.~(\ref{l_4f}) is invariant under the $U(1)$ 
transformation $\psi\to e^{i\phi}\psi$. The $U(1)$ symmetry 
implies that the fermion number 
\be
 N= \int d^3x\,\psi^\dagger \psi
\ee
is conserved. We introduce a chemical potential $\mu$ conjugate 
to the fermion number $N$ and study the partition function
\be 
\label{Z}
 Z(\mu,\beta) = {\rm Tr}\left[e^{-\beta(H-\mu N)}\right].
\ee
Here, $H$ is the Hamiltonian associated with ${\mathcal L}$ and 
$\beta=1/T$ is the inverse temperature. The average number of 
particles for a given chemical potential $\mu$ and temperature 
$T$ is given by $\langle N\rangle =T(\partial \log Z)/(\partial 
\mu)$. At zero temperature the chemical potential is the energy
required to add one particle to the system. 

\begin{figure}[t]
\includegraphics[width=11.0cm]{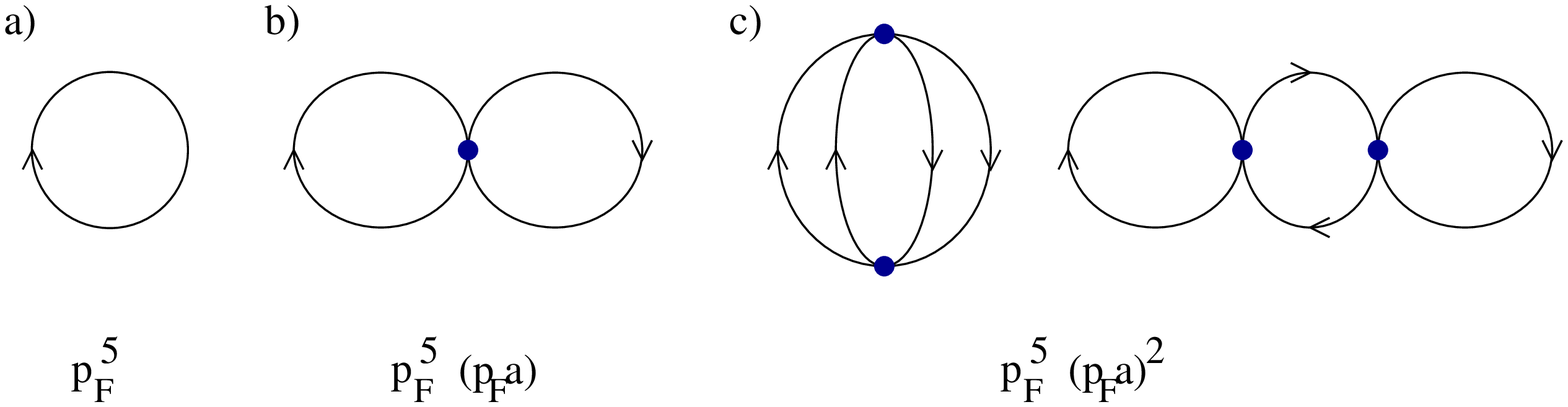}
\caption{\label{fig_fl}
Leading order Feynman diagrams for the ground state 
energy of a dilute gas of fermions interacting via
a short range potential.}
\end{figure}

  We observe that the chemical potential simply shifts the energy 
in the lagrangian. This implies that we have to carefully analyze 
the boundary conditions in the path integral in order to fix the 
pole prescription. The correct Minkowski space propagator is
\be
\label{s_ph} 
S^0_{\alpha\beta}(p) =
 \frac{\delta_{\alpha\beta}}
 {p_0-\epsilon_p+i\delta{\rm sgn}(\epsilon_p)}
 = \delta_{\alpha\beta}\left\{
 \frac{\Theta(p-p_F)}{p_0-\epsilon_p+i\delta}+
 \frac{\Theta(p_F-p)}{p_0-\epsilon_p-i\delta}
  \right\},
\ee
where $\epsilon_p=E_p-\mu$, $E_p=\vec{p}^{\, 2}/(2m)$ and
$\delta\to 0^+$. The quantity $p_F=\sqrt{2m\mu}$ is called 
the Fermi momentum. The two terms in equ.~(\ref{s_ph}) have a simple 
physical interpretation. At finite density and zero temperature all 
states with momenta below the Fermi momentum are occupied, while all 
states above the Fermi momentum are empty. The possible excitation of 
the system are particles above the Fermi surface or holes below the 
Fermi surface, corresponding to the first and second term in 
equ.~(\ref{s_ph}). The particle density is given by
\be
\frac{N}{V} = \int \frac{d^4p}{(2\pi)^4} S^0_{\alpha\alpha}(p)
 \left. e^{ip_0\eta}\right|_{\eta\to 0^+}
 = 2\int \frac{d^3p}{(2\pi)^3}\Theta(p_F-p)
 = \frac{p_F^3}{3\pi^2}.
\ee 
As a first simple application we can compute the energy 
density as a function of the fermion density. For free
fermions, we find 
\be
\label{e0}
{\mathcal E} =  2\int \frac{d^3p}{(2\pi)^3}E_p\Theta(p_F-p)
 = \frac{3}{5}\frac{p_F^2}{2m}\frac{N}{V}.
\ee
We can also compute the corrections to the ground state
energy due to the interaction $\frac{1}{2}C_0(\psi^\dagger
\psi)^2$. The first term is a two-loop diagram with one 
insertion of $C_0$, see Fig.~\ref{fig_fl}. We have
\be 
\label{e1}
{\mathcal E}_1 = C_0\left(\frac{p_F^3}{6\pi^2}\right)^2.
\ee
We should note that equ.~(\ref{e1}) contains two possible 
contractions, called the direct and the exchange term. If the 
fermions have spin $s$ and degeneracy $g=(2s+1)$ then equ.~(\ref{e1}) 
has to be multiplied by a factor $g(g-1)/2$. We also note that 
the sum of the first two terms in the energy density 
can be written 
as 
\be
\label{e_pfa}
\frac{E}{N} = \frac{p_F^2}{2m}\left(
\frac{3}{5} + \frac{2}{3\pi}(p_Fa)+\ldots  \right),
\ee
which shows that the $C_0$ term is the first term in an expansion 
in $p_Fa$, suitable for a dilute, weakly interacting, Fermi gas. The 
expansion in $(p_Fa)$ was carried out to order $(p_Fa)^2$ by Huang, 
Lee and Yang \cite{Lee:1957,Huang:1957}. Since then, the accuracy
was pushed to \cite{Fetter:1971,Hammer:2000xg} $O((p_Fa)^4\log(p_Fa))$.

\begin{figure}[t]
\bc\includegraphics[width=9cm]{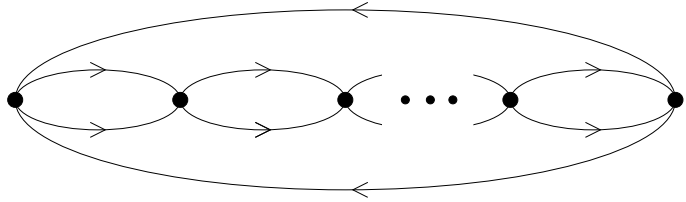}\ec
\caption{\label{fig_lad}
Particle-particle ladder diagrams for the ground state 
energy of a dilute gas of fermions.}
\end{figure}

\subsection{Unitary Limit}
\label{sec_uni}

 The neutron-neutron scattering length is very large, $a_{nn}=-18$ fm, 
and the $(p_Fa)$ expansion is not particularly useful. Indeed, since 
the scattering length is so much larger than all other hadronic
length scales it makes sense to consider the opposite limit and 
take the scattering length to infinity. This means that there 
is a bound state right at threshold and that the low energy cross
section saturates the unitarity bound. If neutron matter is 
dilute then we can also assume that $(p_F r)\ll 1$, where $r$ 
is the range of the potential. In this limit the only energy 
scale in the problem is $p_F^2/(2m)$ and the energy per particle 
is 
\be
\label{xi}
\frac{E}{N}=\xi \frac{3}{5}\frac{p_F^2}{2m},
\ee
where $\xi$ is an unknown parameter. Comparison with equ.~(\ref{e0})
shows that for free fermions $\xi=1$. 

 Neutron matter in the unitary limit is very strongly correlated
and the determination of $\xi$ is a complicated, non-perturbative
problem. However, since $\xi$ is insensitive to the details of 
the interaction the result is the same for any dilute Fermi gas
with a two-body bound state near threshold. It is now possible
to create such a system in the laboratory by trapping cold
fermionic atoms. In these systems the scattering length can 
be controlled using Feshbach resonances induced by an external
magnetic field. A recent experimental analysis yields the
value \cite{Kinast:2005} $\xi\simeq 0.45$.

\begin{figure}[t]
\bc\includegraphics[width=7.0cm]{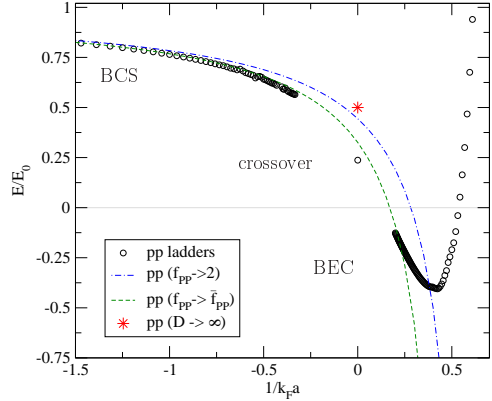}\ec
\caption{\label{fig_bcs_bec}
Total energy of an interacting fermion gas in units
of the energy of a free fermion gas as a function of $(k_Fa)^{-1}$.
The open circles show the result of a numerical calculation of
particle-particle ladder diagrams. The dashed curve shows the 
approximations given in equ.~(\ref{pp_lad}). The star is the 
result of the $d\to\infty$ calculation in the unitary limit.}
\end{figure}

 There have been a number of theoretical attempts to determine
$\xi$. Since the two-body interaction is large it is natural 
to begin with the sum of all two-body diagrams, see Fig.~\ref{fig_lad}.
This sum gives \cite{Schafer:2005kg}
\be
\label{pp_lad}
\frac{E}{N} =\frac{p_F^2}{2M}\left\{ \frac{3}{5} + 
 \frac{2(k_Fa)/(3\pi)}{1-\frac{6}{35\pi}(11-2\log(2))(p_Fa)}\right\}.
\ee
from which we deduce $\xi\simeq 0.32$. This is reasonably close to
the experimental result, but since the system is strongly correlated 
there is no obvious reason to restrict ourselves to two-body ladders. 
We have recently studied the possibility that equ.~(\ref{pp_lad}) 
can be justified as the leading term in an expansion in $1/d$, where 
$d$ is the number of space dimensions \cite{Steele:2000qt,Schafer:2005kg}.
This approach appears promising, but $1/d$ corrections have not been 
worked out yet. Another possibility is to pursue numerical approaches. 
Green function Monte Carlo calculations give $\xi=0.44$, in very good
agreement with the experimental result \cite{Carlson:2003wm}. Several
groups have performed euclidean lattice 
calculations\cite{Chen:2003vy,Wingate:2004wm,Lee:2004qd,Bulgac:2005pj},
similar to the lattice QCD calculations discussed in Sect.~\ref{sec_lqcd}.
These calculations do not suffer from a sign problem and can be 
extended to finite temperature. 

\subsection{Nuclear Matter and Chiral Restoration}
\label{sec_nuc}

 Ordinary nuclei consist of roughly equal numbers of neutrons 
and protons. In order to study heavy nuclei it is useful to 
consider nuclear matter in pure QCD, i.e. ignoring the 
contribution from electrons as well as the Coulomb repulsion 
between protons. As discussed in Sect.~\ref{sec_dense_intro}
nuclear matter saturates at a density $\rho_0\simeq 0.15\, 
{\rm fm}^{-3}$. The binding energy of nuclear matter is 
$B/A\simeq 15$ MeV. Numerical calculations based on realistic 
nucleon-nucleon potentials are successful in reproducing 
these numbers, but we do not understand very well why nuclear
matter saturates and how the saturation density and the binding
energy are related to more fundamental properties of QCD.

 We also do not know very well how to extrapolate the 
equation of state beyond the saturation density. An important 
question is whether we expect chiral symmetry to be 
restored as the density increases. If the density is small 
this question can be studied using the method we employed
in Sect.~\ref{sec_arg}. The quark condensate is given by
\be
\langle\bar{q}q\rangle_\rho = T \frac{\partial}{\partial m_q}
\log Z.
\ee
The partition function of a dilute gas of protons and neutrons is
\be
\log Z=4\int\frac{d^3p}{(2\pi)^3}\log\Big( 1+e^{-(E_N-\mu)/T}\Big).
\ee
The quark mass dependence of the nucleon mass is related to 
$\pi N$ Sigma term $\Sigma_{\pi N}=m_q\partial m_N/\partial m_q$.
We get
\be
\langle\bar{q}q\rangle_\rho = 4\int\frac{d^3p}{(2\pi)^3}
\frac{M_N}{E_N}\Big(\frac{\partial{M_N}}{\partial m_q}\Big)
 \Theta(p_F-|\vec{p}|)
 =\langle\bar{q}q\rangle_0 
\Big\{ 1-\frac{\Sigma_{\pi N}\rho_0}{m_\pi^2 f_\pi^2}
 \Big(\frac{\rho}{\rho_0}\Big)\Big\}.
\ee
The Sigma term can be extracted in pion-nucleon scattering.
Using $\Sigma_{\pi N}\simeq 45$ MeV we find
\be
\langle\bar{q}q\rangle_\rho\simeq\langle\bar{q}q\rangle_0
\Big\{ 1-\frac{1}{3} \Big(\frac{\rho}{\rho_0}\Big)\Big\},
\ee
which indicates that chiral condensate is significantly 
modified already at nuclear matter saturation density.

\subsection{Superfluidity}
\label{sec_bcs}

\begin{figure}[t]
\includegraphics[width=11.0cm]{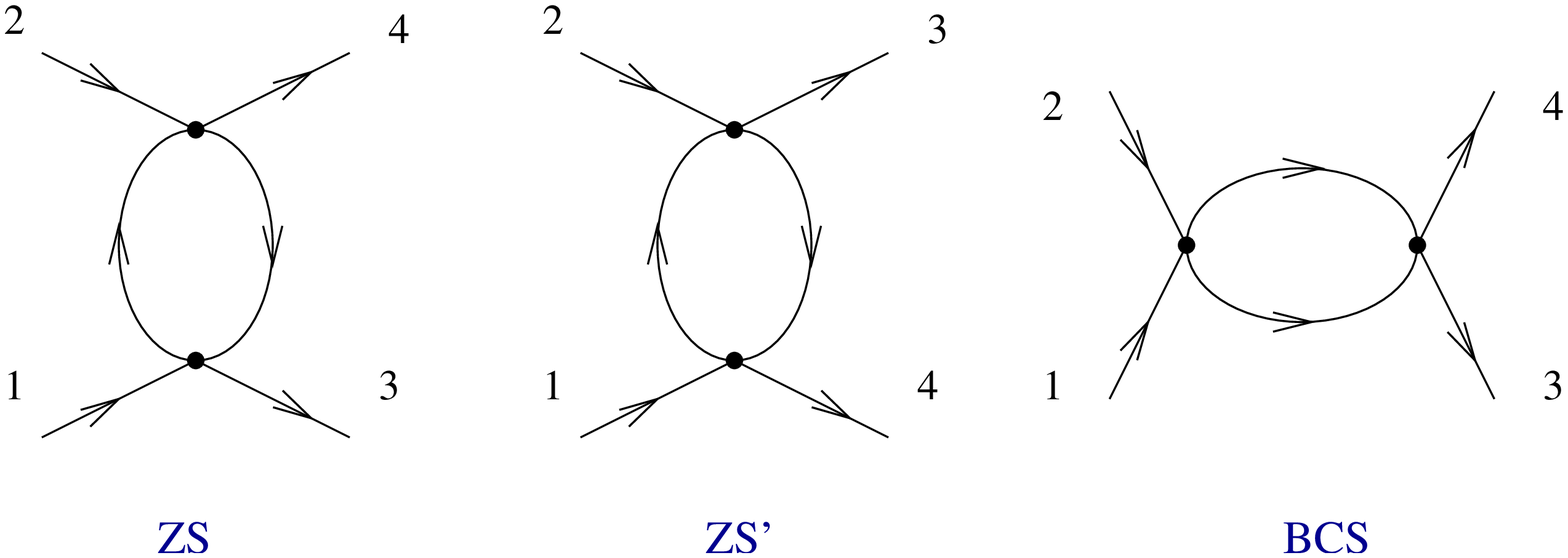}
\caption{\label{fig_bcs}
Second order diagrams that contribute to particle-particle
scattering. The three diagrams are known as ZS (zero sound),
ZS' and BCS (Bardeen-Cooper-Schrieffer) contribution.}
\end{figure}

 One of the most remarkable phenomena that take place in many body 
systems is superfluidity. Superfluidity is related to an instability 
of the Fermi surface in the presence of attractive interactions between 
fermions. Let us consider fermion-fermion scattering in the simple
model introduced in Sect.~\ref{sec_fl}. At leading order the scattering 
amplitude is given by
\be
\label{pp_0}
\Gamma_{\alpha\beta\gamma\delta}(p_1,p_2,p_3,p_4) = 
C_0 \left( \delta_{\alpha\gamma}\delta_{\beta\delta}
 - \delta_{\alpha\delta}\delta_{\beta\gamma} \right).
\ee
At next-to-leading order we find the corrections shown in Fig.~\ref{fig_bcs}. 
A detailed discussion of the role of these corrections can be found in 
\cite{Abrikosov:1963,Shankar:1993pf,Polchinski:1992ed}. The BCS diagram 
is special, because in the case of a spherical Fermi surface it can lead 
to an instability in weak coupling. The main point is that if the 
incoming momenta satisfy $\vec{p}_1\simeq -\vec{p}_2$ then there are 
no kinematic restrictions on the loop momenta. As a consequence, all 
back-to-back pairs can mix and there is an instability even in weak 
coupling. 

 For $\vec{p}_1= -\vec{p}_2$ and $E_1=E_2=E$ the BCS diagram is given by
\bea
\label{diag_bcs}
\Gamma_{\alpha\beta\gamma\delta} &=& 
C_0^2 \left( \delta_{\alpha\gamma}\delta_{\beta\delta}
 - \delta_{\alpha\delta}\delta_{\beta\gamma} \right)
\int \frac{d^4q}{(2\pi)^4} 
 \frac{1}{E+q_0-\epsilon_q+i\delta{\rm sgn}(\epsilon_q)}
 \nonumber \\
 & & \hspace{4cm}
 \frac{1}{E-q_0-\epsilon_q+i\delta{\rm sgn}(\epsilon_q)}.
\eea
The loop integral has an infrared divergence near the Fermi surface 
as $E\to 0$. The scattering amplitude is proportional to 
\be 
\label{cor_bcs}
\Gamma_{\alpha\beta\gamma\delta} =
\left( \delta_{\alpha\gamma}\delta_{\beta\delta}
 - \delta_{\alpha\delta}\delta_{\beta\gamma} \right)
\left\{
C_0 - C_0^2\left(\frac{p_Fm}{2\pi^2}\right)
 \log\left(\frac{E_0}{E}\right) \right\},
\ee
where $E_0$ is an ultraviolet cutoff. Equ.~(\ref{cor_bcs}) can be 
interpreted as an effective energy dependent coupling that satisfies 
the renormalization group equation \cite{Shankar:1993pf,Polchinski:1992ed}
\be 
\label{rge_bcs}
 E\frac{dC_0}{dE} = C_0^2 \left(\frac{p_Fm}{2\pi^2}\right),
\ee
with the solution
\be
\label{rge_sol}
C_0(E) =\frac{C_0(E_0)}{1+NC_0(E_0)\log(E_0/E)},
\ee
where $N=(p_Fm)/(2\pi^2)$ is the density of states. Equ.~(\ref{rge_sol}) 
shows that there are two possible scenarios. If the initial coupling is 
repulsive, $C_0(E_0)>0$, then the renormalization group evolution will 
drive the effective coupling to zero and the Fermi liquid is stable. If, 
on the other hand, the initial coupling is attractive, $C_0(E_0)<0$, then 
the effective coupling grows and reaches a Landau pole at 
\be 
\label{E_lp}
 E_{\it crit} \sim E_0 
    \exp\left(-\frac{1}{N|C_0(E_0)|}\right).
\ee
At the Landau pole the Fermi liquid description has to break down. The 
renormalization group equation does not determine what happens at this 
point, but it seems natural to assume that the strong attractive interaction
will lead to the formation of a fermion pair condensate. The fermion 
condensate $\langle\epsilon^{\alpha\beta}\psi_\alpha\psi_\beta\rangle$ 
signals the breakdown of the $U(1)$ symmetry and leads to a gap $\Delta$ 
in the single particle spectrum. 

 The scale of the gap is determined by the position of the Landau pole, 
$\Delta\sim E_{\it crit}$. A more quantitative estimate of the gap can 
be obtained in the mean field approximation. In the path integral formulation 
the mean field approximation is most easily introduced using the 
Hubbard-Stratonovich trick. For this purpose we first rewrite the 
four-fermion interaction as
\be 
\label{4f_fierz}
\frac{C_0}{2}(\psi^\dagger\psi)^2  = 
\frac{C_0}{4} \left\{
 (\psi^\dagger\sigma_2\psi^\dagger)
 (\psi\sigma_2\psi) 
+(\psi^\dagger\sigma_2\vec{\sigma}\psi^\dagger)
 (\psi\vec{\sigma}\sigma_2\psi)\right\},
\ee 
where we have used the Fierz identity $2\delta^{\alpha\beta}
\delta^{\gamma\rho} = \delta^{\alpha\rho}\delta^{\gamma\beta}+
(\vec{\sigma})^{\alpha\rho}(\vec{\sigma})^{\gamma\beta}$. Note that 
the second term in equ.~(\ref{4f_fierz}) vanishes because $(\sigma_2
\vec{\sigma})$ is a symmetric matrix. We now introduce a factor of 
unity into the path integral
\be
1 = \frac{1}{Z_\Delta}\int D\Delta 
\exp\left(\frac{\Delta^*\Delta}{C_0}\right),
\ee
where we assume that $C_0<0$. We can eliminate the four-fermion 
term in the lagrangian by a shift in the integration variable $\Delta$. 
The action is now quadratic in the fermion fields, but it involves 
a Majorana mass term $\psi\sigma_2\Delta \psi+h.c$. The
Majorana mass terms can be handled using the Nambu-Gorkov 
method. We introduce the bispinor $\Psi=(\psi,\psi^\dagger
\sigma_2)$ and write the fermionic action as
\be
\label{s_ng}
{S} = \frac{1}{2}\int\frac{d^4p}{(2\pi)^4}
 \Psi^\dagger
 \left(\begin{array}{cc}
     p_0-\epsilon_p  & \Delta \\
     \Delta^* & p_0+\epsilon_p
 \end{array}\right) \Psi.
\ee
Since the fermion action is quadratic we can integrate
the fermion out and obtain the effective lagrangian
\be
\label{s_ng_eff}
L= \frac{1}{2}{\rm Tr}\left[\log\left(
 G_0^{-1}G\right)\right]+\frac{1}{C_0}|\Delta|^2,
\ee
where $G$ is the fermion propagator
\be
\label{ng_prop}
 G(p) = \frac{1}{p_0^2-\epsilon_p^2-|\Delta|^2}
 \left(\begin{array}{cc}
     p_0+\epsilon_p  & \Delta^* \\
     \Delta & p_0-\epsilon_p
 \end{array}\right).
\ee
The diagonal and off-diagonal components of $G(p)$ are 
sometimes referred to as normal and anomalous propagators. 
Note that we have not yet made any approximation. We have 
converted the fermionic path integral to a bosonic one, albeit 
with a very non-local action. The mean field approximation 
corresponds to evaluating the bosonic path integral using 
the saddle point method. Physically, this approximation
means that the order parameter does not fluctuate. 
Formally, the mean field approximation can be 
justified in the large $N$ limit, where $N$ is the
number of fermion fields. The saddle point equation
for $\Delta$ gives the gap equation
\be
\Delta = |C_0|\int\frac{d^4p}{(2\pi)^4} 
 \frac{\Delta}{p_0^2-\epsilon^2_p-\Delta^2}.
\ee
Performing the $p_0$ integration we find
\be
\label{4f_gap}
1 = \frac{|C_0|}{2}\int\frac{d^3p}{(2\pi)^3} 
 \frac{1}{\sqrt{\epsilon^2_p+\Delta^2}}.
\ee
Since $\epsilon_p=E_p-\mu$ the integral in equ.~(\ref{4f_gap}) 
has an infrared divergence on the Fermi surface $|\vec{p}| \sim 
p_F$. As a result, the gap equation has a non-trivial solution even 
if the coupling is arbitrarily small. The magnitude of the gap is 
$\Delta\sim \Lambda \exp(-1/(|C_0|N))$ where $\Lambda$ is a cutoff 
that regularizes the integral in equ.~(\ref{4f_gap}) in the ultraviolet. 
If we treat equ.~(\ref{l_4f}) as a low energy effective field theory 
we should be able to eliminate the unphysical dependence of the gap 
on the ultraviolet cutoff, and express the gap in terms of a physical 
observable. At low density this can be achieved by observing that 
the gap equation has the same UV behavior as the Lipmann-Schwinger 
equation that determines the scattering length at zero density
\be
\label{bubble}
\frac{mC_0}{4\pi a} - 1 = \frac{C_0}{2} 
\int\frac{d^3p}{(2\pi)^3}\frac{1}{E_P}.
\ee
Combining equs.~(\ref{4f_gap}) and (\ref{bubble}) we can derive an 
UV finite gap equation that depends only on the scattering length, 
\be
-\frac{m}{4\pi a} = 
\frac{1}{2}\int\frac{d^3p}{(2\pi)^3} \Big\{
 \frac{1}{\sqrt{\epsilon^2_p+\Delta^2}}
 -\frac{1}{E_p}\Big\}.
\ee 
Solving for the $\Delta$ we find \cite{Papenbrock:1998wb,Khodel:1996}
\be
\label{gap_lowd}
\Delta = \frac{8E_f}{e^2}\exp\left(-\frac{\pi}{2p_F|a|}\right).
\ee
Higher order corrections reduce the pre-exponent in this 
result by a factor $(4e)^{1/3}\simeq 2.2$ \cite{Gorkov:1961}.
Like the perturbative calculation of the energy per particle 
this result is not very useful for neutron matter, since the 
scattering length is very large. Taking higher order corrections
into account, Equ.~(\ref{gap_lowd}) suggests that $\Delta \sim 
0.49 E_f$ as $p_F|a|\to\infty$. Surprisingly, this estimate 
agrees very well with numerical calculations \cite{Chang:2004sj}. 
The gap is also quite sensitive to the effective range of the
interaction. Calculations based on potential models give gaps 
on the order of 2 MeV at nuclear matter density.

\subsection{Landau-Ginzburg theory}
\label{sec_lg}

 In neutron stars there is not only pairing between neutrons 
but also pairing between protons. Since protons carry charge 
this implies that the material is not only a superfluid but 
also a superconductor. Superconductors have many interesting 
properties which can be understood from the symmetries 
involved. We will consider a system of protons coupled to a 
$U(1)$ gauge field $A_\mu$. The order parameter $\Phi=\langle 
\epsilon^{\alpha\beta}\psi_\alpha\psi_\beta\rangle$ breaks 
$U(1)$ invariance. Consider a gauge transformation 
\be 
A_\mu\to A_\mu +\partial_\mu\Lambda .
\ee
The order parameter transforms as
\be 
\Phi \to \exp(2ie\Lambda)\Phi.
\ee
The breaking of gauge invariance is responsible for most of the 
unusual properties of superconductors \cite{Anderson:1984,Weinberg:1995}.
This can be seen by constructing the low energy effective action 
of a superconductor. For this purpose we write the order parameter
in terms of its modulus and phase
\be 
\Phi(x) = \exp(2ie\phi(x)) \tilde\Phi(x).
\ee
The field $\phi$ corresponds to the Goldstone mode. Under a gauge
transformation $\phi(x)\to\phi(x)+\Lambda(x)$. 
Gauge invariance restricts the form of the effective Lagrange
function 
\be 
\label{L_sc}
 L = -\frac{1}{4}\int d^3x\, F_{\mu\nu}F_{\mu\nu}
 + L_s (A_\mu-\partial_\mu\phi).
\ee
There is a large amount of information we can extract even 
without knowing the explicit form of $L_s$. Stability implies
that $A_\mu=\partial_\mu\phi$ corresponds to a minimum of the 
energy. This means that up to boundary effects the gauge 
potential is a total divergence and that the magnetic field
has to vanish. This phenomenon is known as the Meissner
effect. 

 Equ.~(\ref{L_sc}) also implies that a superconductor
has zero resistance. The equations of motion relate
the time dependence of the Goldstone boson field to 
the potential, 
\be
\label{phidot}
\dot\phi(x)=-V(x).
\ee 
The electric current is related to the gradient of the 
Goldstone boson field. Equ.~(\ref{phidot}) shows that the 
time dependence of the current is proportional to the 
gradient of the potential. In order to have a static 
current the gradient of the potential has to be constant
throughout the sample, and the resistance is zero. 

 In order to study the properties of a superconductor in more 
detail we have to specify $L_s$. For this purpose we assume 
that the system is time-independent, that the spatial 
gradients are small, and that the order parameter is small. 
In this case we can write
\be 
\label{l_lg}
 L_s = \int d^3x\, \left\{
-\frac{1}{2}\left|\left(\nabla-2ie\vec{A}\right)\Phi\right|^2
 +\frac{1}{2}m^2_H\left(\Phi^*\Phi\right)^2
 -\frac{1}{4}g\left(\Phi^*\Phi\right)^4 + \ldots \right\},
\ee
where $m_H$ and $g$ are unknown parameters that depend
on the temperature. Equ.~(\ref{l_lg}) is known as the 
Landau-Ginzburg effective action. Strictly speaking, the 
assumption that the order parameter is small can only be 
justified in the vicinity of a second order phase transition. 
Nevertheless, the Landau-Ginzburg description is instructive 
even in the regime where $t=(T-T_c)/T_c$ is not small. It is 
useful to decompose $\Phi=\rho\exp(2ie\phi)$. For constant
fields the effective potential, 
\be
\label{v_lg}
V(\rho)=-\frac{1}{2}m_H^2\rho^2 +\frac{1}{4}g\rho^4 ,
\ee
is independent of $\phi$. The minimum is at $\rho_0^2=m_H^2/g$ 
and the energy density at the minimum is given by ${E}= 
-m_H^4/(4g)$. This shows that the two parameters $m_H$ and
$g$ can be related to the expectation value of $\Phi$ and
the condensation energy. We also observe that the phase
transition is characterized by $m_H(T_c)=0$. 

 In terms of $\phi$ and $\rho$ the Landau-Ginzburg action 
is given by
\be 
 L_s = \int d^3x\, \left\{
-2e^2\rho^2 \left(\vec\nabla\phi-\vec{A}\right)^2
 +\frac{1}{2}m_H^2\rho^2 -\frac{1}{4}g\rho^4
 -\frac{1}{2}\left(\nabla\rho\right)^2
\right\}.
\ee
The equations of motion for $\vec{A}$ and $\rho$ 
are given by
\bea 
\label{b_lg}
\vec\nabla\times \vec{B} &=& 
 4e^2\rho^2 \left(\nabla\phi -\vec{A}\right), \\
\label{rho_lg}
 \nabla^2 \rho &=& 
 -m_H^2\rho^2 + g\rho^3 + 4e^2 \rho 
 \left( \vec\nabla\phi-\vec{A}\right) .
\eea
Equ.~(\ref{b_lg}) implies that $\nabla^2\vec{B} = -4e^2\rho^2\vec{B}$. 
This means that an external magnetic field $\vec{B}$ decays over a 
characteristic distance $\lambda=1/(2e\rho)$. Equ.~(\ref{rho_lg}) 
gives $\nabla^2\rho = -m_H^2\rho+\ldots$. As a consequence, variations 
in the order parameter relax over a length scale given by $\xi=1/m_H$. 
The two parameters $\lambda$ and $\xi$ are known as the penetration 
depth and the coherence length. 

 The relative size of $\lambda$ and $\xi$ has important consequences 
for the properties of superconductors. In a type II superconductor 
$\xi<\lambda$. In this case magnetic flux can penetrate the system 
in the form of vortex lines. At the core of a vortex the order parameter
vanishes, $\rho=0$. In a type II material the core is much smaller 
than the region over which the magnetic field goes to zero. The 
magnetic flux is given by
\be
\int_A\vec{B}\cdot\vec{S} =
\oint_{\partial A} \vec{A}\cdot d\vec{l} = 
\oint_{\partial A} \vec{\nabla}\phi \cdot d\vec{l} =
\frac{n\pi\hbar}{e} ,
\ee
and quantized in units of $\pi\hbar/e$. In a type II superconductor 
magnetic vortices repel each other and form a regular lattice known 
as the Abrikosov lattice. In a type I material, on the other hand, 
vortices are not stable and magnetic fields can only penetrate
the sample if superconductivity is destroyed. 

\section{QCD at high density}
\label{sec_dqcd}
\subsection{Color superconductivity}
\label{sec_csc}

 In Sect.~\ref{sec_dense_intro} we introduced a few simple arguments 
concerning the phase diagram of QCD in the $\mu-T$ plane. These arguments 
are summarized in Fig.~\ref{fig_phase_1}. The basic idea is that large 
baryon density is just like high temperature: there is a large scale in 
the problem, the effective coupling is weak, and the system is described, 
to a good approximation, as a weakly interacting quark liquid. We expect,
in particular, that quarks are deconfined and that chiral symmetry is
restored. 

 We also showed, however, that systems at finite density, as exemplified
by nuclear matter, have a very rich phase diagram. We saw, in particular, 
that the BCS instability will lead to pair condensation whenever there 
is an attractive fermion-fermion interaction, even if the interaction
is weak. At very large density, the attraction is provided by 
one-gluon exchange between quarks in a color anti-symmetric $\bar 3$ 
state. High density quark matter is therefore expected to behave as a 
color superconductor \cite{Frau_78,Barrois:1977xd,Bar_79,Bailin:1984bm}.

  Color superconductivity is described by a pair condensate of the form
\be
\label{csc}
\Phi = \langle \psi^TC\Gamma_D\lambda_C\tau_F\psi\rangle.
\ee
Here, $C$ is the charge conjugation matrix, and $\Gamma_D,
\lambda_C,\tau_F$ are Dirac, color, and flavor matrices. 
Except in the case of only two colors, the order parameter
cannot be a color singlet. Color superconductivity is 
therefore characterized by the breakdown of color gauge 
invariance. This statement has to be interpreted in the 
sense of Sect.~\ref{sec_lg}. Gluons acquire a mass due
to the (Meissner-Anderson) Higgs mechanism. 

\begin{figure}[t]
\bc\includegraphics[width=12.0cm]{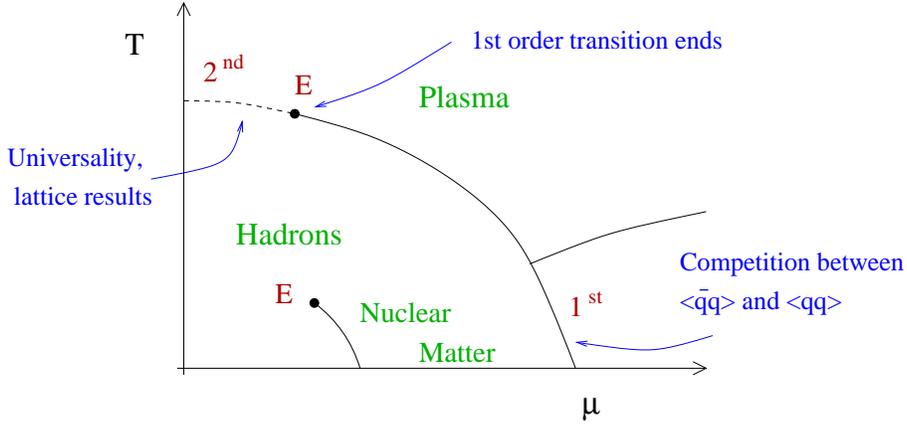}\ec
\caption{\label{fig_phase_2}
First revision of the phase diagram of hadronic matter.
This figure shows the phase diagram of strongly interacting
matter obtained from a mean field treatment of chiral symmetry 
breaking and color superconductivity in QCD with two flavors.}
\end{figure}

A rough estimate of the critical density for the transition
from chiral symmetry breaking to color superconductivity, the 
superconducting gap and the transition temperature is provided 
by schematic four-fermion models \cite{Alford:1998zt,Rapp:1998zu}.
Typical models are based on the instanton interaction 
\be 
\label{l_I}
{\mathcal L} = G_{I}\left\{
 (\bar\psi\tau^-_\alpha\psi)^2 + 
 (\bar\psi\gamma_5\tau^-_\alpha\psi)^2 
 \right\},
\ee
or a schematic one-gluon exchange interaction 
\be 
\label{l_OGE}
{\mathcal L} = G_{OGE}\left(\bar{\psi}\gamma_\mu
 \frac{\lambda^a}{2}\psi\right)^2 . 
\ee
Here $\tau^-_\alpha=(\vec{\tau},i)$ is an isospin matrix and
$\lambda^a$ are the color Gell-Mann matrices. The strength 
of the four-fermion interaction is typically tuned to reproduce 
the magnitude of the chiral condensate and the pion decay 
constant at zero temperature and density. In the mean field 
approximation the effective quark mass associated with chiral 
symmetry breaking is determined by a gap equation of the type 
\be 
\label{m_gap}
 M_Q = G_M  \int^\Lambda\frac{d^3p}{(2\pi)^3} 
  \frac{M_Q}{\sqrt{{\vec{p}}^{\,2}+M_Q^2}}
   \left(1-n_F(E_p)\right),
\ee
where $G_M$ is the effective coupling in the quark-anti-quark
channel, $\Lambda$ is a cutoff, and $n_F(E)$ is the Fermi 
distribution. Both the instanton interaction 
and the one-gluon exchange interaction are attractive in the color 
anti-triplet scalar diquark channel $\epsilon^{abc}(\psi^b C\gamma_5
\psi^c)$. A pure one-gluon exchange interaction leads to a 
degeneracy between scalar and pseudoscalar diquark condensation, 
but instantons are repulsive in the pseudoscalar diquark 
channel. The gap equation in the scalar diquark channel is 
\be
\label{d_gap}
 \Delta = \frac{G_D}{2} \int^\Lambda\frac{d^3p}{(2\pi)^3} 
  \frac{\Delta}{\sqrt{(|\vec{p}|-p_F)^2+\Delta^2}},
\ee
where we have neglected terms that do not have a singularity
on the Fermi surface $|\vec{p}|=p_F$. In the case of a 
four-fermion interaction with the quantum numbers of one-gluon 
exchange $G_D=G_M/(N_c-1)$. The same result holds for instanton 
effects. In order to determine the correct ground state we have 
to compare the condensation energy in the chiral symmetry broken 
and diquark condensed phases. We have ${E} \sim f_\pi^2M_Q^2$ in 
the $(\bar{q}q)$ condensed phase and ${E}\sim p_F^2\Delta^2/(2\pi^2)$ 
in the $(qq)$ condensed phase. 

\begin{figure}[t]
\bc\includegraphics[width=9.0cm]{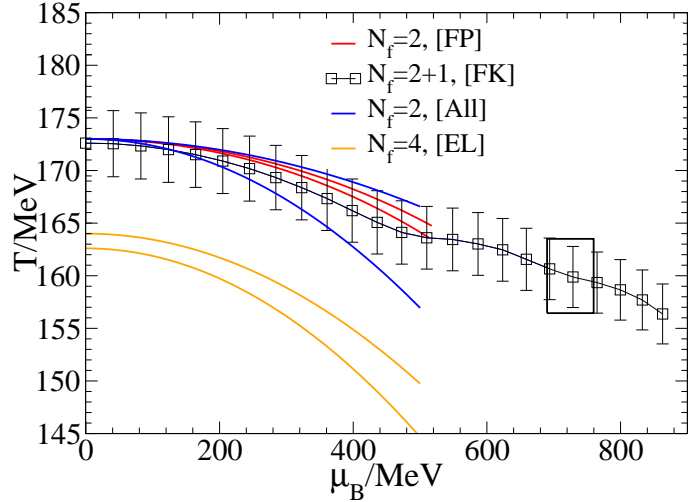}\ec
\caption{\label{fig_tri}
Location of the (pseudo) critical line and tri-critical point (box) 
as measured in different simulations, de Forcrand and Philipsen
(2002) [FP], Fodor and Katz (2002) [FK], Allton et al (2002) 
[All], D'Ellia and Lombardo (2002) [EL]. Figure from de Forcrand
and Philipsen (2003).}
\end{figure}

 At zero temperature and density both equs.~(\ref{m_gap}) and 
(\ref{d_gap}) only have non-trivial solutions if the coupling 
exceeds a critical value. Since $G_M>G_D$ we have $M_Q>\Delta$ and 
the energetically preferred solution corresponds to chiral symmetry 
breaking. If the density increases Pauli-Blocking in equ.~(\ref{m_gap})
becomes important and the effective quark mass decreases. The diquark 
gap equation behaves very differently. Equ.~(\ref{d_gap}) has an infrared 
singularity on the Fermi surface, $p=p_F$, and this singularity is 
multiplied by a finite density of states, $N=p_F^2/(2\pi)^2$. As a 
consequence, there is a non-trivial solution even if the coupling is 
weak. The gap grows with density until the Fermi momentum becomes on 
the order of the cutoff. For realistic values of the parameters we 
find a first order transition for remarkably small values of the quark 
chemical potential, $\mu_Q\simeq 300$ MeV. The gap in the diquark 
condensed phase is $\Delta\sim 100$ MeV and the critical temperature 
is $T_c\sim 50$ MeV.

 In the same model the finite temperature phase transition at 
zero baryon density is found to be of second order. This result 
is in agreement with universality arguments \cite{Pisarski:ms}
and lattice results. If the transition at finite density and zero 
temperature is indeed of first order then the first order transition 
at zero baryon density has to end in a tri-critical point 
\cite{Barducci:1989wi,Berges:1998,Halasz:1998}. 
 The tri-critical point is quite remarkable, because it remains a true 
critical point even if the quark masses are not zero. A non-zero quark
mass turns the second order $T\neq 0$ transition into a smooth crossover, 
but the first order $\mu\neq 0$ transition persists. It is hard to 
predict where exactly the tri-critical point is located in the phase 
diagram. Recent lattice calculations suggest that the tri-critical 
point is sufficiently close to the finite temperature axis so that 
its location can be determined on the lattice, see Fig.~\ref{fig_tri}.
It may also be possible to locate the critical point experimentally. 
Heavy ion collisions at relativistic energies produce matter under the 
right conditions and experimental signatures of the tri-critical point 
have been suggested in \cite{Stephanov:1998}.

 A schematic phase diagram is shown in Fig.~\ref{fig_phase_2}. We 
should emphasize that this phase diagram is based on simplified models 
and that there is no proof that the transition from nuclear matter to 
quark matter along the $T=0$ line occurs via a single first order 
transition. Chiral symmetry breaking and color superconductivity 
represent two competing forms of order, and it seems unlikely that 
the two phases are separated by a second order transition. However, 
since color superconductivity modifies the spectrum near the Fermi 
surface, whereas chiral symmetry breaking operates near the surface 
of the Dirac sea, it is not clear that the two phases cannot coexist. 
Indeed, there are models in which a phase coexistence region appears 
\cite{Kitazawa:2002bc}.

\subsection{Phase structure in weak coupling}
\label{sec_phases}
\subsubsection{QCD with two flavors}
\label{sec_nf2}

  In this section we shall discuss how to use weak coupling methods 
in order to explore the phases of dense quark matter. We begin with 
what is usually considered to be the simplest case, quark matter with 
two degenerate flavors, up and down. Renormalization group arguments 
suggest \cite{Evans:1999ek,Schafer:1999na}, and explicit calculations 
confirm \cite{Brown:1999yd,Schafer:2000tw}, that whenever possible quark 
pairs condense in an $s$-wave state. This means that the spin wave function 
of the pair is anti-symmetric. Since the color wave function is also 
anti-symmetric, the Pauli principle requires the flavor wave function 
to be anti-symmetric too. This essentially determines the structure of 
the order parameter \cite{Alford:1998zt,Rapp:1998zu}
\be
\label{2sc}
\Phi^a  = \langle \epsilon^{abc}\psi^b C\gamma_5 \tau_2\psi^c
 \rangle.
\ee
This order parameter breaks the color $SU(3)\to SU(2)$ and
leads to a gap for up and down quarks with two out of the 
three colors. Chiral and isospin symmetry remain unbroken. 

\begin{figure}[t]
\bc\includegraphics[width=11.0cm]{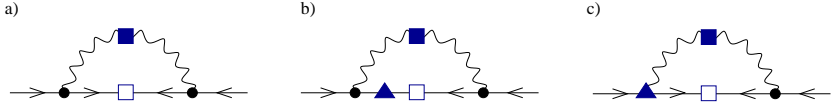}\ec
\caption{\label{fig_gap}
Fig.~a) shows the leading contribution to the Dyson-Schwinger 
(gap) equation in QCD at finite density. The open square denotes an 
anomalous self energy (gap) insertion and the solid square is a gluon 
self energy insertion. Figs.~b) and c) show  quark self energy insertions 
and vertex corrections.}
\end{figure}

  We can calculate the magnitude of the gap and the condensation energy 
using weak coupling methods. In weak coupling the gap is determined by 
ladder diagrams with the one gluon exchange interaction. These diagrams 
can be summed using the gap equation 
\cite{Son:1999uk,Schafer:1999jg,Pisarski:2000tv,Hong:2000fh,Brown:1999aq}
\bea
\label{eliash}
\Delta(p_4) &=& \frac{g^2}{12\pi^2} \int dq_4\int d\cos\theta\,
 \left(\frac{\frac{3}{2}-\frac{1}{2}\cos\theta}
            {1-\cos\theta+G/(2\mu^2)}\right. \\
 & & \hspace{3cm}\left.    +\frac{\frac{1}{2}+\frac{1}{2}\cos\theta}
            {1-\cos\theta+F/(2\mu^2)} \right)
 \frac{\Delta(q_4)}{\sqrt{q_4^2+\Delta(q_4)^2}}. \nonumber
\eea
Here, $\Delta(p_4)$ is the frequency dependent gap, $g$ is the QCD 
coupling constant and $G$ and $F$ are the self energies of magnetic 
and electric gluons. This gap equation is very similar to the BCS gap 
equation equ.~(\ref{d_gap}) obtained in four-fermion models. The terms 
in the curly brackets arise from the magnetic and electric components 
of the gluon propagator. The numerators are the on-shell matrix elements 
${M}_{ii,00}=[\bar{u}_h(p_1)\gamma_{i,0}u_h(p_3)][\bar{u}_h(p_2)\gamma_{i,0}
u_h(p_4)]$ for the scattering of back-to-back fermions on the Fermi surface. 
The scattering angle is $\cos\theta=\hat{p}_1\cdot\hat{p}_3$. In the case 
of a spin zero order parameter, the helicity $h$ of all fermions is the 
same, see \cite{Schafer:1999jg} for more detail. 

 The main difference between equ.~(\ref{eliash}) and the BCS gap equation 
(\ref{d_gap}) is that because the gluon is massless, the gap equation 
contains a collinear divergence for $\cos\theta\sim 1$. In a dense medium 
the collinear divergence is regularized by the gluon self energy. For 
$\vec{q}\to 0$ and to leading order in perturbation theory we have
\be
\label{pi_qcd}
 F = 2m^2, \hspace{1cm}
 G = \frac{\pi}{2}m^2\frac{q_4}{|\vec{q}|},
\ee
with $m^2=N_fg^2\mu^2/(4\pi^2)$. In the electric part, $m_D^2=2m^2$ is 
the familiar Debye screening mass. In the magnetic part, there is no 
screening of static modes, but non-static modes are modes are dynamically 
screened due to Landau damping. This is completely analogous to the 
situation at finite temperature \cite{Manuel:1995td}, see 
Sect.~\ref{sec_pqcd}.

 For small energies dynamic screening of magnetic modes is much weaker 
than Debye screening of electric modes. As a consequence, perturbative 
color superconductivity is dominated by magnetic gluon exchanges. Using 
equ.~(\ref{pi_qcd}) we can perform the angular integral in equ.~(\ref{eliash})
and find
\be
\label{eliash_mel}
\Delta(p_4) = \frac{g^2}{18\pi^2} \int dq_4
 \log\left(\frac{b\mu}{\sqrt{|p_4^2-q_4^2|}}\right)
    \frac{\Delta(q_4)}{\sqrt{q_4^2+\Delta(q_4)^2}},
\ee
with $b=256\pi^4(2/N_f)^{5/2}g^{-5}$. We can now see why it was important 
to keep the frequency dependence of the gap. Because the collinear divergence 
is regulated by dynamic screening, the gap equation depends on $p_4$
even if the frequency is small. We can also see that the gap scales 
as $\exp(-c/g)$. The collinear divergence leads to a gap equation with 
a double-log behavior. Qualitatively
\be
\label{dlog}
 \Delta \sim \frac{g^2}{18\pi^2}\Delta 
 \left[\log\left(\frac{\mu}{\Delta}\right)\right]^2,
\ee
from which we conclude that $\Delta\sim\exp(-c/g)$. Equ.~(\ref{dlog}) 
is not sufficiently accurate to determine the correct value of the 
constant $c$. A more detailed analysis shows that the gap on the 
Fermi surface is given by
\be
\label{gap_oge}
\Delta_0 \simeq 512\pi^4(2/N_f)^{5/2}b'\mu g^{-5}
   \exp\left(-\frac{3\pi^2}{\sqrt{2}g}\right).
\ee
The factor $b'$ is related to non-Fermi liquid effects, see 
\cite{Brown:2000eh,Ipp:2003cj,Schafer:2004zf}. In perturbation 
theory $b'=\exp(-(\pi^2+4)(N_c-1)/16)$ 
\cite{Brown:1999aq,Wang:2001aq,Schafer:2003jn}. The condensation 
energy is given by
\be
\epsilon = -N_d \Delta_0^2\left(\frac{\mu^2}{4\pi^2}\right),
\ee
where $N_d=4$ is the number of condensed species. In the mean field 
approximation the critical temperature is $T_c/\Delta_0 =e^\gamma/
\pi\simeq 0.56$, as in standard BCS theory \cite{Pisarski:2000tv}. 
Fluctuations of the gauge field drive the transition first order
\cite{Bailin:1984bm,Matsuura:2003md}. Surprisingly, gauge field
fluctuations increase the critical temperature as compared to the 
BCS result \cite{Giannakis:2004xt}.

 For chemical potentials $\mu<1$ GeV, the coupling constant is not 
small and the applicability of perturbation theory is in doubt. If 
we ignore this problem and extrapolate the perturbative calculation 
to densities $\rho\simeq 5\rho_0$ we find gaps on the order of 10's 
of MeV. However, perturbative estimates also show that instanton
effects cannot be neglected for $\mu<1$ GeV, and that instantons
increase the gap \cite{Schafer:2004yx}.

 We note that the 2SC phase defined by equ.~(\ref{2sc}) has two gapless 
fermions and an unbroken $SU(2)$ gauge group. The gapless fermions are 
singlets under the unbroken $SU(2)$. As a consequence, we expect the 
$SU(2)$ gauge group to become non-perturbative. An estimate of the 
$SU(2)$ confinement scale was given in \cite{Rischke:2000cn}. We also 
note that even though the Copper pairs carry electric charge the $U(1)$
of electromagnetism is not broken. The generator of this symmetry is a 
linear combination of the original electric charge operator and the 
diagonal color charges. Under this symmetry the gapless fermions carry 
the charges of the proton and neutron. Possible pairing between the 
gapless fermions was discussed in \cite{Alford:1998zt,Alford:2002xx}.
 
\begin{figure}[t]
\includegraphics[width=9.0cm]{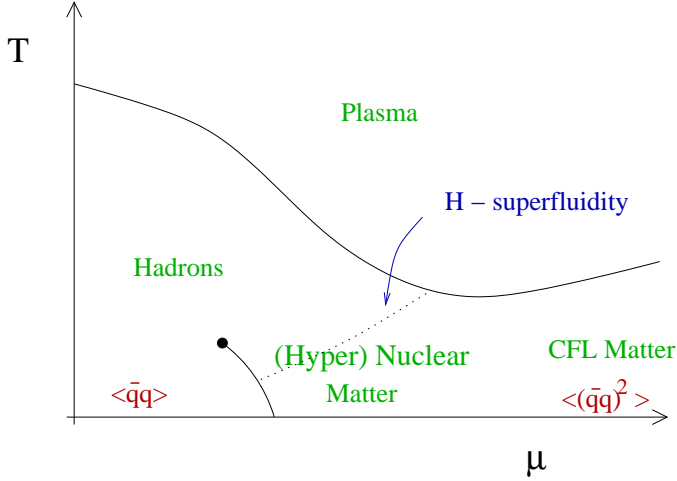}
\caption{\label{fig_phase_3}
Conjectured phase diagram of $N_f=3$ hadronic matter in the 
limit of exact flavor symmetry.}
\end{figure}

\subsubsection{QCD with three flavors: Color-Flavor-Locking}
\label{sec_cfl} 

 If quark matter is formed at densities several times nuclear matter 
density we expect the quark chemical potential to be larger than the 
strange quark mass. We therefore have to determine the structure of the 
superfluid order parameter for three quark flavors. We begin with the 
idealized situation of three degenerate flavors. From the arguments 
given in the last section we expect the order parameter to be 
color-flavor matrix of the form
\be
\label{order}
  \Phi^{ab}_{ij}=
  \langle \psi^a_i C\gamma_5\psi^b_j\rangle.
\ee
The structure of this matrix can be determined by extremizing 
the grand canonical potential. We find \cite{Schafer:1999fe,Evans:1999at}
\be
\label{cfl}
\Delta^{ab}_{ij} = 
 \Delta_A (\delta_i^a\delta_j^b-\delta_i^b\delta_j^a)
+\Delta_S (\delta_i^a\delta_j^b+\delta_i^b\delta_j^a),
\ee
which describes the color-flavor locked (CFL) phase proposed by Alford, 
Rajagopal, and Wilczek \cite{Alford:1999mk}. In the weak coupling limit 
$\Delta_S \ll\Delta_A$ and $\Delta_A=2^{-1/3}\Delta_0$ where $\Delta_0$ 
is the gap in the 2SC phase, equ.~(\ref{gap_oge}) \cite{Schafer:1999fe}. 
In the CFL phase both color and flavor symmetry are completely broken. 
There are eight combinations of color and flavor symmetries that generate 
unbroken global symmetries. The unbroken symmetries are
\be
\psi^a_{L,i}\to (U^*)^{ab}U_{ij}\psi^b_{Lj},
\hspace{1cm}
\psi^a_{R,i}\to (U^*)^{ab}U_{ij}\psi^b_{Rj},
\ee
for $U\in SU(3)_V$. The symmetry breaking pattern is 
\be
\label{sym_3}
SU(3)_L\times SU(3)_R\times U(1)_V\to SU(3)_V .
\ee
We observe that color-flavor-locking implies that chiral symmetry is 
broken. The mechanism for chiral symmetry breaking is quite unusual. The 
primary order parameter $\langle \psi^a_{Li}C\Delta^{ab}_{ij}\psi^b_{Lj}
\rangle=-\langle \psi^a_{Ri}C\Delta^{ab}_{ij}\psi^b_{Rj}\rangle$ involves 
no coupling between left and right handed fermions. In the CFL phase both 
left and right handed flavor are locked to color, and because of the 
vectorial coupling of the gluon left handed flavor is effectively locked 
to right handed flavor. Chiral symmetry breaking also implies that 
$\langle \bar{\psi}\psi\rangle$ has a non-zero expectation value
\cite{Schafer:2002ty}. In the CFL phase $\langle \bar{\psi}\psi
\rangle^2\ll \langle(\bar{\psi}\psi)^2 \rangle$. Another measure of 
chiral symmetry breaking is provided by the pion decay constant. 
We will see that in the weak coupling limit $f_\pi^2$ is proportional 
to the density of states on the Fermi surface.

 The symmetry breaking pattern $SU(3)_L\times SU(3)_R \to SU(3)_V$ 
in the CFL phase is identical to the symmetry breaking pattern in QCD at 
low density. The spectrum of excitations in the color-flavor-locked (CFL) 
phase also looks remarkably like the spectrum of QCD at low density 
\cite{Schafer:1999ef}. The excitations can be classified according to 
their quantum numbers under the unbroken $SU(3)$, and by their electric 
charge. The modified charge operator that generates a true symmetry of 
the CFL phase is given by a linear combination of the original charge 
operator $Q_{em}$ and the color hypercharge operator $Q={\rm diag}(-2/3,
-2/3,1/3)$. Also, baryon number is only broken modulo 2/3, which means 
that one can still distinguish baryons from mesons. We find that the 
CFL phase contains an octet of Goldstone bosons associated with chiral 
symmetry breaking, an octet of vector mesons, an octet and a singlet of 
baryons, and a singlet Goldstone boson related to superfluidity. All of 
these states have integer charges.  

  With the exception of the $U(1)$ Goldstone boson, these states exactly 
match the quantum numbers of the lowest lying multiplets in QCD at low 
density. In addition to that, the presence of the $U(1)$ Goldstone boson 
can also be understood. The $U(1)$ order parameter is $\langle (uds)(uds)
\rangle$. This order parameter has the quantum numbers of a $0^+$ $(\Lambda
\Lambda)$ pair condensate. In $N_f=3$ QCD, this is the most symmetric two 
nucleon channel, and a very likely candidate for superfluidity in nuclear 
matter at low to moderate density. We conclude that in QCD with three
degenerate light flavors, there is no fundamental difference between the 
high and low density phases. This implies that a low density hyper-nuclear 
phase and the high density quark phase might be continuously connected, 
without an intervening phase transition. A conjectured phase diagram is 
shown in Fig.~\ref{fig_phase_3}. 

\subsection{The role of the strange quark mass}
\label{sec_ms} 
 
  At baryon densities relevant to astrophysical objects dis\-tor\-tions 
of the pure CFL state due to non-zero quark masses cannot be neglected 
\cite{Alford:1999pa,Schafer:1999pb}. The most important effect of a 
non-zero strange quark mass is that the light and strange quark 
Fermi momenta will no longer be equal. When the mismatch is much 
smaller than the gap one calculates assuming degenerate quarks, we 
might expect that it has very little consequence, since at this 
level the original particle and hole states near the Fermi surface 
are mixed up anyway. On the other hand, when the mismatch is much 
larger than the nominal gap,we might expect that the ordering one 
would obtain for degenerate quarks is disrupted, and that to a first 
approximation one can treat the light and heavy quark dynamics separately.

\begin{figure}[t]
\bc\includegraphics[width=10cm]{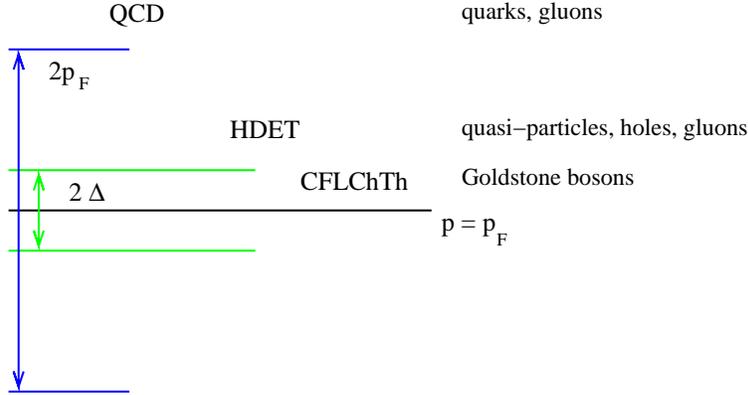}\ec
\caption{\label{fig_eft}
Hierarchy of effective field theories in the CFL phase.}
\end{figure}

 This argument is qualitatively right, but the correct picture 
turns out to be much more complicated, and much more interesting. 
If the strange quark mass is taken into account microscopic 
calculations based on the Dyson-Schwinger equation become much 
more complicated, because there are many more gap parameters, 
and maintaining electric neutrality and color gauge invariance 
is difficult \cite{Steiner:2002gx,Alford:2002kj,Neumann:2002jm}. 
However, since chiral symmetry is broken in the CFL phase we know 
that the dependence on the quark masses is constrained by chiral 
symmetry. It is therefore natural to study the problem using 
effective field theories. In practice we will employ a two-step 
procedure, see Fig.~\ref{fig_eft}. In the first step we match the 
microscopic theory, QCD, to an effective field theory of quasi-particles 
and holes in the vicinity of the Fermi surface. In the second step we 
match this theory to an effective chiral theory for the CFL phase.

\subsubsection{High density effective theory}
\label{sec_hdet}

 The QCD Lagrangian in the presence of a chemical potential is given by
\be
\label{qcd}
 {\mathcal L} = \bar\psi \left( i\Dslash +\mu\gamma_0 \right)\psi
 -\bar\psi_L M\psi_R - \bar\psi_R M^\dagger \psi_L 
 -\frac{1}{4}G^a_{\mu\nu}G^a_{\mu\nu},
\ee
where $D_\mu=\partial_\mu+igA_\mu$ is the covariant derivative, $M$ is 
the mass matrix and $\mu$ is the baryon chemical potential. If the baryon 
density is very large perturbative QCD calculations can be further 
simplified. The main observation is that the relevant degrees of 
freedom are particle and hole excitations in the vicinity of the 
Fermi surface. We shall describe these excitations in terms of the 
field $\psi_+(\vec{v},x)$, where $\vec{v}$ is the Fermi velocity. 
At tree level, the quark field $\psi$ can be decomposed as $\psi=
\psi_++\psi_-$ where $\psi_\pm=\frac{1}{2}(1\pm\vec{\alpha}\cdot\hat{v})
\psi$. Note that $(1\pm\vec{\alpha}\cdot\hat{v})/2$ is a projector
on states with positive/negative energy. To leading order in $1/p_F$ 
we can eliminate the field $\psi_-$ using its equation of motion. The 
lagrangian for the $\psi_+$ field is given by 
\cite{Hong:2000tn,Hong:2000ru,Beane:2000ms}
\be
\label{hdet}
{\mathcal L} =  \psi_{+}^\dagger (iv\cdot D) \psi_{+}
  - \frac{ \Delta}{2}\left(\psi_{+}^{ai} C \psi_{+}^{bj}
 \left(\delta_{ai}\delta_{bj}-
           \delta_{aj}\delta_{bi} \right) 
           + {\rm h.c.} \right) + \ldots ,
\ee
with $v_\mu=(1,\vec{v})$ and $i,j,\ldots$ and $a,b,\ldots$ denote 
flavor and color indices. The magnitude of the gap $\Delta$ is 
determined order by order in perturbation theory from the requirement 
that the thermodynamic potential is stationary with respect to $\Delta$. 
With the gap term included the perturbative expansion is well defined.

\begin{figure}[t]
\bc\includegraphics[width=7.5cm]{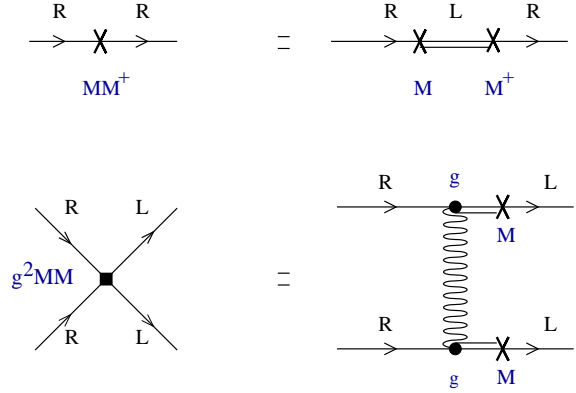}\ec
\caption{\label{fig_hdet_m}
Mass terms in the high density effective theory. The first 
diagram shows a $O(MM^\dagger)$ term that arises from integrating 
out the $\psi_-$ field in the QCD lagrangian. The second
diagram shows a $O(M^2)$ four-fermion operator which arises from 
integrating out $\psi_-$ and hard gluon exchanges.}
\end{figure}

  The effective theory contains an infinite set of operators that
have additional derivatives or more powers of $\psi_+$. These
operators are suppressed by inverse powers of the Fermi momentum. 
Here, we will only consider operators that contain quark masses. To 
leading order in $1/p_F$ there is only one operator in the high density 
effective theory
\be 
\label{m_kin}
{\mathcal L} = -\frac{1}{2p_F} \left( \psi_{L+}^\dagger MM^\dagger \psi_{L+}
 + \psi_{R+}^\dagger M^\dagger M\psi_{R+} \right).
\ee
This term arises from expanding the kinetic energy of a massive
fermion around $p=p_F$. At $O(1/p_F^2)$ we find four-fermion 
operators that contain two powers of the quark mass. The 
coefficients of these operators are obtained by computing 
chirality violating quark-quark scattering amplitudes for 
quasi-particles near the Fermi surface \cite{Schafer:2001za}, 
see Fig.~\ref{fig_hdet_m}. At leading order in $1/p_F$ these
amplitudes are independent of the scattering angle and can be 
represented as local four-fermion operators
\be
\label{hdet_m}
 {\mathcal L} = \frac{g^2}{8p_F^4}
 \left( ({\psi^A_L}^\dagger C{\psi^B_L}^\dagger)
        (\psi^C_R C \psi^D_R) \Gamma^{ABCD} +
        ({\psi^A_L}^\dagger \psi^B_L) 
        ({\psi^C_R}^\dagger \psi^D_R) \tilde{\Gamma}^{ACBD} \right).
\ee
There are two additional terms with $(L\leftrightarrow R)$ and
$(M\leftrightarrow M^\dagger)$. We have introduced the CFL 
eigenstates $\psi^A$ defined by $\psi^a_i=\psi^A (\lambda^A)_{ai}
/\sqrt{2}$, $A=0,\ldots,8$. The tensor $\Gamma$ is defined by
\bea 
 \Gamma^{ABCD} &=& \frac{1}{8}\Big\{ {\rm Tr} \left[ 
    \lambda^A M(\lambda^D)^T \lambda^B M (\lambda^C)^T\right] 
  \nonumber \\
 & & \hspace{1cm}\mbox{}
   -\frac{1}{3} {\rm Tr} \left[
    \lambda^A M(\lambda^D)^T \right]
    {\rm Tr} \left[
    \lambda^B M (\lambda^C)^T\right] \Big\}.
\eea
The explicit expression for $\tilde\Gamma$ is given 
in \cite{Schafer:2001za}, but we will not need here. 

\subsubsection{CFL chiral theory}
\label{sec_CFLchi}

 For excitation energies smaller than the gap the only relevant 
degrees of freedom are the Goldstone modes associated with the 
breaking of chiral symmetry and baryon number, see Fig.~\ref{fig_eft}. 
Since the pattern of chiral symmetry breaking is identical to 
the one at $T=\mu=0$ the effective lagrangian has the same 
structure as chiral perturbation theory. The main difference is
that Lorentz-invariance is broken and only rotational invariance
is a good symmetry. The effective lagrangian for the Goldstone 
modes is \cite{Casalbuoni:1999wu}
\bea
\label{l_cheft}
{\mathcal L}_{eff} &=& \frac{f_\pi^2}{4} {\rm Tr}\left[
 \nabla_0\Sigma\nabla_0\Sigma^\dagger - v_\pi^2
 \partial_i\Sigma\partial_i\Sigma^\dagger \right] 
 +\left[ B {\rm Tr}(M\Sigma^\dagger) + h.c. \right] 
    \nonumber \\ 
 & & \hspace*{0cm}\mbox{} 
     +\left[ A_1{\rm Tr}(M\Sigma^\dagger)
                        {\rm Tr} (M\Sigma^\dagger) 
     + A_2{\rm Tr}(M\Sigma^\dagger M\Sigma^\dagger) \right.
 \nonumber \\[0.1cm] 
  & &   \hspace*{0.5cm}\mbox{}\left. 
     + A_3{\rm Tr}(M\Sigma^\dagger){\rm Tr} (M^\dagger\Sigma)
         + h.c. \right]+\ldots . 
\eea
Here $\Sigma=\exp(i\phi^a\lambda^a/f_\pi)$ is the chiral field,
$f_\pi$ is the pion decay constant and $M$ is a complex mass
matrix. The chiral field and the mass matrix transform as
$\Sigma\to L\Sigma R^\dagger$ and  $M\to LMR^\dagger$ under 
chiral transformations $(L,R)\in SU(3)_L\times SU(3)_R$. We 
have suppressed the singlet fields associated with the breaking 
of the exact $U(1)_V$ and approximate $U(1)_A$ symmetries. 

 At low density the coefficients $f_\pi$, $B,A_i,\ldots$ are 
non-perturbative quantities that have to extracted from 
experiment or measured on the lattice. At large density, on
the other hand, the chiral coefficients can be calculated in 
perturbative QCD. The leading order terms are \cite{Son:1999cm} 
\be
\label{cfl_fpi}
f_\pi^2 = \frac{21-8\log(2)}{18} 
  \left(\frac{p_F^2}{2\pi^2} \right), 
\hspace{0.5cm} v_\pi^2=\frac{1}{3}.
\ee
Mass terms are determined by the operators studied in the 
previous section. We observe that both equ.~(\ref{m_kin})
and (\ref{hdet_m}) are quadratic in $M$. This implies that $B=0$
in perturbative QCD. $B$ receives non-perturbative contributions
from instantons, but these effects are small if the density is
large \cite{Schafer:2002ty}. 

  We observe that $X_L=MM^\dagger/(2p_F)$ and $X_R=M^\dagger M/
(2p_F)$ in equ.~(\ref{m_kin}) act as effective chemical potentials 
for left and right-handed fermions, respectively. Formally, the 
effective lagrangian has an $SU(3)_L\times SU(3)_R$ gauge 
symmetry under which $X_{L,R}$ transform as the temporal components
of non-abelian gauge fields. We can implement this approximate gauge 
symmetry in the CFL chiral theory by promoting time derivatives
to covariant derivatives \cite{Bedaque:2001je}, 
\be
\label{mueff}
 \nabla_0\Sigma = \partial_0 \Sigma 
 + i \left(\frac{M M^\dagger}{2p_F}\right)\Sigma
 - i \Sigma\left(\frac{ M^\dagger M}{2p_F}\right) .
\ee
The four-fermion operator in equ.~(\ref{hdet_m}) contributes to 
the coefficients $A_i$.We find \cite{Son:1999cm,Schafer:2001za}
\be
 A_1= -A_2 = \frac{3\Delta^2}{4\pi^2}, 
\hspace{1cm} A_3 = 0.
\ee

 We can now summarize the structure of the chiral expansion in the
CFL phase. The effective lagrangian has the form 
\be
{\mathcal L}\sim f_\pi^2\Delta^2 \left(\frac{\partial_0}{\Delta}\right)^k
 \left(\frac{\vec{\partial}}{\Delta}\right)^l
 \left(\frac{MM^\dagger}{p_F\Delta}\right)^m
 \left(\frac{MM}{p_F^2}\right)^n
  \big(\Sigma\big)^o\big(\Sigma^\dagger\big)^p.
\ee
Loop graphs in the effective theory are suppressed by powers of 
$\partial/(4\pi f_\pi)$. Since the pion decay constant scales as $f_\pi
\sim p_F$ Goldstone boson loops are suppressed compared to higher 
order contact terms. We also note that the quark mass expansion 
is controlled by $m^2/(p_F\Delta)$, as expected from the arguments
presented in Sect.~\ref{sec_ms}.

\begin{figure}[t]
\bc\includegraphics[width=9.5cm]{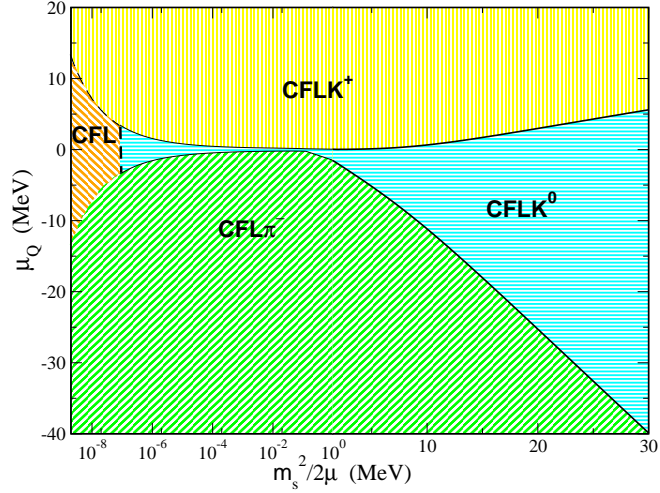}\ec
\caption{\label{fig_kcond}
This figure shows the phase structure of CFL matter as 
a function of the strange quark mass $m_s$ and the lepton
chemical potential $\mu_Q$, from Kaplan and Reddy (2001).}
\end{figure}

\subsubsection{Kaon condensation}
\label{sec_kcond}

 Using the chiral effective lagrangian we can now determine 
the dependence of the order parameter on the quark masses. We will 
focus on the physically relevant case $m_s>m_u=m_d$. Because
the main expansion parameter is $m_s^2/(p_F\Delta)$ increasing 
the quark mass is roughly equivalent to lowering the density. 
The effective potential for the order parameter is 
\be
\label{v_eff}
V_{eff} = \frac{f_\pi^2}{2} {\rm Tr}\left[
 X_L\Sigma X_R\Sigma^\dagger \right] 
     + A_1\left[ \left({\rm Tr}(M\Sigma^\dagger)\right)^2 
     - {\rm Tr}(M\Sigma^\dagger M\Sigma^\dagger) \right].
\ee
If the strange quark mass is small then the minimum of the 
effective potential is $\Sigma=1$. However, when the strange 
quark mass exceeds a certain critical value it becomes favorable
to rotate the order parameter in the kaon direction. The physical
reason is that the system tries to reduce its strangeness content 
by forming a kaon condensate. Consider the ansatz $\Sigma = \exp
(i\alpha\lambda_4)$. The vacuum energy is 
\be 
\label{k0+_V}
 V(\alpha) = -f_\pi^2 \left( \frac{1}{2}\left(\frac{m_s^2-m^2}{2p_F}
   \right)^2\sin(\alpha)^2 + (m_{K}^0)^2(\cos(\alpha)-1)
   \right),
\ee
where $(m_K^0)^2= (4A_1/f_\pi^2)m(m+m_s)$. Minimizing the vacuum 
energy we obtain $\alpha=0$ if $\mu_s<m_K^0$ and $\cos(\alpha)
=(m_K^0)^2/\mu_s^2$ if $\mu_s >m_K^0$. Here, we have defined 
$\mu_s=m_s^2/(2p_F)$. Using the perturbative result for $A_1$
the critical strange quark mass is 
\be
\label{ms_crit}
\left. m_s \right|_{crit}= 3.03\cdot  m_d^{1/3}\Delta^{2/3}.
\ee
Using $\Delta\simeq 50$ MeV we get $m_s(crit)\simeq 70$ MeV. This 
result suggests that strange quark matter at densities $\rho \sim
(5-10)\rho_0$ is in a kaon condensed phase. The kaon condensate 
breaks $SU(2)_I\times U(1)_Y$ to $U(1)_Q$. The phase structure 
as a function of the strange quark mass and non-zero lepton chemical 
potentials was studied by Kaplan and Reddy \cite{Kaplan:2001qk}, 
see Fig.~\ref{fig_kcond}. We observe that if the lepton chemical 
potential is non-zero charged kaon and pion condensates are also 
possible. It was also shown that there is a range of light quark 
masses in which simultaneous kaon and eta condensation takes 
place \cite{Kryjevski:2004cw}.

\begin{figure}[t]
\bc\includegraphics[width=9.5cm]{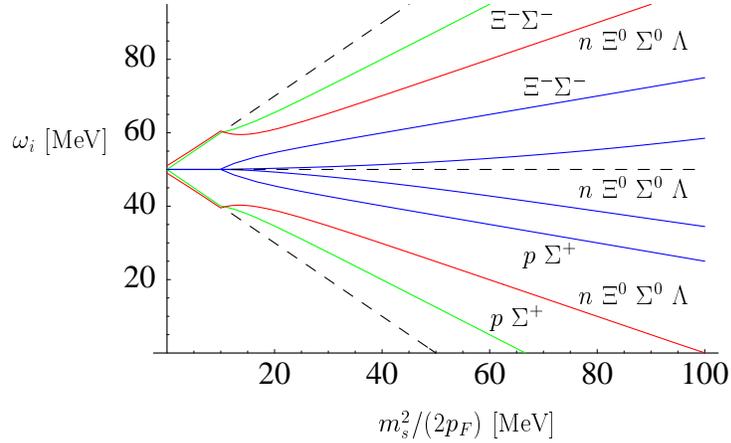}\ec
\caption{\label{fig_cfl_spec}
This figure shows the fermion spectrum in the CFL phase. For 
$m_s=0$ there are eight fermions with gap $\Delta$ and one
fermion with gap $2\Delta$ (not shown). Without kaon condensation
gapless fermion modes appear at $\mu_s=\Delta$ (dashed lines).
With kaon condensation gapless modes appear at $\mu_s=4\Delta/3$.}
\end{figure}

\section{Fermions in the CFL phase}
\label{sec_gCFL}

 So far we have only studied Goldstone modes in the CFL phase.
However, as the strange quark mass is increased it is possible
that some of the fermion modes become light or even gapless
\cite{Alford:2003fq}. In order to study this question we 
have to include fermions in the effective field theory. 
The effective lagrangian for fermions in the CFL phase
is \cite{Kryjevski:2004jw,Kryjevski:2004kt}
\bea 
\label{l_bar}
{\mathcal L} &=&  
 {\rm Tr}\left(N^\dagger iv^\mu D_\mu N\right) 
 - D{\rm Tr} \left(N^\dagger v^\mu\gamma_5 
               \left\{ {\mathcal A}_\mu,N\right\}\right)
 - F{\rm Tr} \left(N^\dagger v^\mu\gamma_5 
               \left[ {\mathcal A}_\mu,N\right]\right)
  \nonumber \\
 & &  \mbox{} + \frac{\Delta}{2} \left\{ 
     \left( {\rm Tr}\left(N_LN_L \right) 
   - \left[ {\rm Tr}\left(N_L\right)\right]^2 \right)  
   - (L\leftrightarrow R) + h.c.  \right\}.
\eea
$N_{L,R}$ are left and right handed baryon fields in the 
adjoint representation of flavor $SU(3)$. The baryon fields 
originate from quark-hadron complementarity as explained
in Sect.~\ref{sec_cfl}. We can think of $N$ as being composed 
of a quark and a diquark field, $N_L \sim q_L\langle q_L q_L
\rangle$. The covariant derivative of the nucleon field is given 
by $D_\mu N=\partial_\mu N +i[{\mathcal V}_\mu,N]$. The vector 
and axial-vector currents are 
\be
 {\mathcal V}_\mu = -\frac{i}{2}\left\{ 
  \xi \partial_\mu\xi^\dagger +  \xi^\dagger \partial_\mu \xi 
  \right\}, \hspace{1cm}
{\mathcal A}_\mu = -\frac{i}{2} \xi\left(\nabla_\mu 
    \Sigma^\dagger\right) \xi , 
\ee
where $\xi$ is defined by $\xi^2=\Sigma$. It follows that $\xi$ 
transforms as $\xi\to L\xi U(x)^\dagger=U(x)\xi R^\dagger$ with 
$U(x)\in SU(3)_V$. For pure $SU(3)$ flavor transformations $L=R=V$ 
we have $U(x)=V$. $F$ and $D$ are low energy constants that 
determine the baryon axial coupling. In perturbative QCD we
find $D=F=1/2$.

 Mass terms can be introduced as in Sect.~\ref{sec_CFLchi}.
The $(X_L,X_R)$ covariant derivative of the nucleon field is  
\bea
\label{V_X}
 D_0N     &=& \partial_0 N+i[\Gamma_0,N], \\
 \Gamma_0 &=& -\frac{i}{2}\left\{ 
  \xi \left(\partial_0+ iX_R\right)\xi^\dagger + 
  \xi^\dagger \left(\partial_0+iX_L\right) \xi 
  \right\}, \nonumber 
\eea
where $X_L=MM^\dagger/(2p_F)$ and $X_R=M^\dagger M/(2p_F)$ as before. 
The spectrum of fermion is shown in Fig.~\ref{fig_cfl_spec}. A gapless
fermion mode appears at $\mu_s=4\Delta/3$. In the vicinity of this 
point the homogeneous CFL phase becomes unstable 
\cite{Huang:2004bg,Casalbuoni:2004tb}. In the effective field theory
this manifests itself as an instability with respect to the generation
of a non-zero current \cite{Kryjevski:2005qq,Schafer:2005ym}. From 
the effective lagrangian equ.~(\ref{l_cheft}) we see that a meson
current has energy ${\mathcal E}\sim f_\pi^2 j^2$. This is not the
end of the story, however, because a meson current also modifies
the fermion dispersion relation. The energy of the lowest mode
in the background of a hypercharge current $j_K$ is given by 
\be
\label{disp_ax}
\omega_l = \Delta +\frac{l^2}{2\Delta}-\frac{3}{4}
  \mu_s -\frac{1}{4}\vec{v}\cdot\vec{j}_K,
\ee
where $l$ is the momentum relative to the Fermi surface. We observe 
that the current lowers the energy of the fermions on part of the Fermi 
surface. When these states become gapless the total energy is lowered
and the system can become unstable. The new ground state is a 
$p$-wave meson condensate in which the non-zero meson current is 
balanced by a backflow of gapless fermions. At even larger values
of $\mu_s$ this state may evolve into an inhomogeneous superconductor
of the type considered by Larkin, Ovchinnikov, Fulde and Ferrell
\cite{Larkin:1964,Fulde:1964}, see \cite{Casalbuoni:2005zp}.

\begin{figure}[t]
\includegraphics[width=11.0cm]{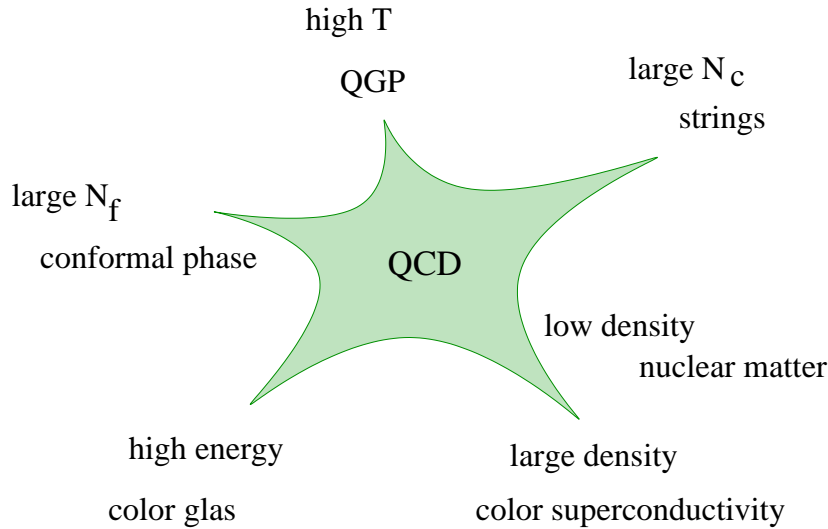}
\caption{\label{fig_toe}
The many facets of QCD.}
\end{figure}

\section{Outlook}
\label{sec_sum}

 Figure \ref{fig_toe} indicates that there are many interesting 
connections that we have not been able to explore in these
lectures. There has been a lot of progress in connecting the 
large $N_c$ world to a theory of strings, and this connection
also sheds some light on the behavior of a strongly coupled 
QCD plasma. The transport properties of the strongly coupled
plasma are probably quite similar to the transport behavior
of the strongly correlated neutron fluid, and this system 
is related to cold trapped fermionic atoms near a Feshbach
resonance. A lot of progress has been made in understanding 
hot and dense pre-equilibrium states, and these states share
some of the properties of equilibrium phases. Many more 
surprising connections are likely to energy in the future.
  
 Acknowledgments: I would like to thank the organizers of the 
HUGS summer school, especially Jose Goity, for their hospitality. 
The original lectures contained a summary of the experimental 
work at RHIC and CERN. This material is not included in the 
write-up, but the slides are still available on my website.
The second half of the lectures is an abridged and updated version 
of the NPSS lectures \cite{Schafer:2003vz}. Fig.~\ref{fig_toe} 
was inspired by R.~Brower. This work was supported in part by 
a US DOE grant DE-FG-88ER40388.


\end{document}